# Electrodynamics of carbon nanotubes with non-local surface conductivity

Tomer Berghaus[1,**], Touvia Miloh[1], Oded Gottlieb[2] and Gregory Ya. Slepyan[3*]

[1] *School of Mechanical Engineering, Tel Aviv University, Tel Aviv 69978, Israel*

[2] *Faculty of Mechanical Engineering, Technion—Israel Institute of Technology, Haifa 32000, Israel*

[3] *School of Electrical Engineering, Tel Aviv University, Tel Aviv 69978, Israel*

A new framework that can be utilized for the electrodynamics of carbon nanotubes (CNTs) with non-local surface conductivity (spatial dispersion) is presented. The model of non-local conductivity is developed on the basis of the Kubo technique applied to the Dirac equation for pseudospins. As a result, the effective boundary conditions for the electromagnetic (EM) field on a CNT surface are formulated. The dispersion relation for the eigenmodes of an infinitely long CNT is obtained and analyzed. It is shown that due to nonlocality, a new type of eigenmodes are created that disappear in the local conductivity limit. These eigenmodes should be properly accounted for in the correct formulation of the CNT end conditions for the surface current, which are manifested in the EM-field scattering problem.  Additional boundary conditions that consider nonlocality effects are also formulated based on the exact solution obtained for the surface current by means of using the Wiener-Hopf (WH) technique for a semi-infinite CNT. The scattering pattern of the EM-field is simulated by a finite-length model of a CNT, using a numerically solved integral equation for the surface current density and its approximate analytical solution. Thus, the scattering field of a CNT prevailing in the wide frequency range from THz to infrared light is analytically solved and analyzed. The newly obtained results are then utilized for determining the optical forces exerted on a CNT of finite length. Potential applications for the design of nanoantennas and other electronic devices, including pointing out some future directions, are also discussed.

## I. INTRODUCTION

Recent progress in nanotechnology has paved the way for the synthesis of artificial materials (metamaterials) with exotic electromagnetic properties, garnering significant attention at both theoretical and experimental levels. Among the most promising classes of materials for industrial and commercial applications are carbon-based nanomaterials, including graphene nano-ribbons and carbon nanotubes (CNTs) [1]-[5]. The electromagnetism of CNTs and graphene continues to be a fertile ground for catalyzing breakthroughs that are likely to extend far beyond current expectations. Technological advancements have enabled their applications in quantum computing [6], quantum informatics and simulations [7], nanoelectronics [2], and biomedicine [8],[9] (including biological nanomotors [9]). One of the most notable areas of research and commercial development is the rapid advancement in terahertz (THz) science and technology

*gregory_slepyan@yahoo.com

*t.berghaus@gmail.com

[9]. The electronic and optical properties of CNTs have been actively studied since the late 1990s. Fundamental phenomena such as the THz conductivity peak [10],[11] and plasmonic resonance [11], [12] have been discovered and extensively reviewed from both theoretical and experimental perspectives.

A detailed understanding of CNT electrodynamics is becoming increasingly critical for various applications, including interconnects [13], [14], nanoantennas [12], [15]–[18], and rectennas [19]. Advances in fabrication have enabled precise control over CNT properties through variations in spatial configuration (e.g., bent CNTs [20]) and by filling their internal cavities with fullerenes, quantum dots [8], color centers [21], polymers [22], and host materials such as conducting linear chains of sulfur [23]. These innovations necessitate the development of new theoretical models to account for the previously neglected physical effects.

One such effect relates to nonlocality of the optical properties, a fundamental aspect of light-matter interaction in condensed matter systems [24]– [27], also known as spatial dispersion. This phenomenon is characterized by a medium's response (e.g., polarization, electric current) at a given point depending not only on position but also on neighboring values of the generalized force (electric or magnetic field). Early research in this area mainly focused on crystal optics [24], superconductivity [25], the anomalous skin effect in normal metals [25], and plasmas [26], driven by the potential to manipulate optical properties through the emergence of novel wave modes. Following a period of low activity, the field has experienced resurgence due to breakthroughs in artificial material fabrication. This revival includes metamaterials [27], [28], metasurfaces [29], mesoscopic metal surfaces [30], materials with near-zero [31] or negative refractive index [32],[33], topological materials [34–36], and graphene-based systems [37]–[41]. First-principle theory confirms that spatial nonlocality is a necessary condition for achieving negative refraction [33].

Early theoretical models were developed for a three-dimensional and spatially homogeneous medium. In these cases, nonlocality expressed in the coordinate space was represented using integral operators with convolution-type kernels. Simplification was achieved by moving to the momentum space and employing Fourier transforms, where spatial dispersion appeared as a dependence of the constitutive parameters (e.g., permittivity, conductivity) on wave momentum. More recent research has shifted toward low-dimensional systems, where boundary effects become significant and complicate the momentum-space approaches. In such cases, the integral kernel is no longer of a convolution form, and a new general operator emerges, whose characterization requires boundary-dependent models that vary in complexity and rigorous theoretical framework.

A central concept in this context is the role of additional boundary conditions (ABCs) which are particularly prominent in the field of crystal optics [24]. It is now recognized that when spatial dispersion is included, ABCs are indeed indispensable. The necessity of including ABCs, has been debated historically (e.g., considered a "historical mistake" in crystal optics [42], a claim which was later refuted [43]– [45]). Nonlocality raises the order of the governing differential equations, often introducing new wave solutions, which in turn require additional boundary conditions. These new waves, such as longitudinal plasma waves [26], further confirm the need for ABCs. As shown in [24], ABCs are sufficient to solve problems involving the reflection and refraction of ordinary waves at the crystal-vacuum interface. However, if additional wave modes propagate in the crystal, the problem becomes underdetermined without supplementary boundary conditions.

The present study advances the emerging field of nonlocal electrodynamics by investigating the impact of spatial dispersion on the optical properties of CNTs. Applying first-principles approaches to these types of problems suggests new fundamental challenges, among them is the difficulty of bridging between condensed matter physics and quantum electrodynamics [46-48]. As an example, one may note the application of quantum electrodynamics and the Kubo technique for the analysis of graphene's nonlocal electrical conductivity [48]. The obtained results point to a disagreement between the two approaches, which initiated the discussion in [28]. However, these contradictions were found only for the transverse component of the conductivity. Due to the quasi-two-dimensional structure of CNTs, their conductivity is primarily manifested through the longitudinal component. Therefore, our formulation is free from these contradictions and can be analytically solved using the above-mentioned techniques. In what follows, we apply the general susceptibility theory combined with the Kubo formalism [49], [50]. The longitudinal electric field serves as the generalized force, and the longitudinal current as the corresponding response. Given the significance of finite-length effects, the analysis is conducted in real space rather than momentum space. Starting with a general integral formulation of nonlocal conductivity, a simplified differential relation is subsequently derived between the surface current density and the longitudinal electric field.

A complete solution incorporating the ABC problem, necessitates specific assumptions regarding the shape (geometry) and termination of the CNT. For instance, in the case of a semi-infinite CNT with perfectly reflecting ends, the Wiener–Hopf (WH) technique is employed. This approach has been previously utilized in electrodynamics, acoustics, and mechanics [51]– [53], and was recently employed in the study of carbon-based nanostructures, including semi-infinite CNTs [54] and twisted bilayer graphene [55]. Its principal advantage lies in enabling exact analytical solutions to rather complex diffraction problems for arbitrary geometries and wavelength regimes. In the context of nonlocal electrodynamics, it also provides a robust framework for deriving the ABCs required to accurately describe wave scattering in CNTs characterized in terms of the tensorial nonlocal conductivity. Our study focuses on achiral CNTs, both metallic and semiconducting. Nonlocal effects and finite-length (edge) effects are considered concurrently.

Our key methodology is enforcing the effective physical boundary conditions [56], wherein a smooth, homogeneous cylinder with the same radius as the real CNT is introduced. The proper boundary conditions for the electromagnetic field are imposed on the cylinder's surface in order to ensure that the resulting field distribution matches that of the actual CNT beyond a short distance from the lattice. The discontinuity in the tangential magnetic field across the CNT surface corresponds to the surface current density, while the tangential electric field remains continuous. To solve Maxwell's equations for a finite-length CNT, we use an integral equation approach in a similar manner to methods developed for macroscopic wire antennas [57]–[62]. These formulations assume that the CNT radius is considered small compared to the wavelength. Numerical challenges often emerge from the use of approximate kernels, which may lead to ill-posed integral equations characterized by nonsingular kernels [63], [64]. These issues are generally manifested by the appearance of a pronounced numerical oscillations, particularly when a large number of basis functions are employed [63], [64]. To address these instability matters, effective regularization strategies—such as those introduced, for example, by Hanson et al. [63]—are indispensable. Furthermore, the Leontovich–Levin equation [61], [62] offers a widely adopted approximate analytical solution, serving as a valuable tool in scenarios where exact formulations are computationally prohibitive.

Unlike macroscopic wire antennas, which assume perfect conductivity, CNTs require modified boundary conditions [65]. Nevertheless, in Maxwellian frameworks, material properties usually enter via relatively simple boundary conditions. All formulations of integral equations for local electromagnetic properties have been adapted for CNTs [66]– [72] and applied to various problems, including CNT nanoantennas [16],[65],[67], spontaneous emission and the Purcell effect [68], wave scattering in crossed CNTs [70], and CNT arrays [72]. These integral equations support both numerical and analytical solutions. Yet, selecting the optimal formulation and regularization remains a non-trivial task. In this work, we adopt the Pocklington-like equation with an exact kernel [63] to ensure numerical accuracy.

The structure of the paper is as follows: Section II addresses the decomposition of the CNT's nonlocal conductivity using the Kubo formalism. Section III develops the theory of eigenmodes in CNTs with spatial dispersion, using the effective boundary conditions. In Section IV, we derive the end conditions (ABCs) using the exact WH solution for a semi-infinite model. Section V, the core of the paper, discusses the integral equation approach for electromagnetic wave scattering by finite-length CNTs with nonlocal conductivity. Section VI presents numerical results and discusses experimental implications. Section VII applies the results to the analysis of optomechanical forces in CNTs and conclusions and outcome are finally outlined in Section VIII.

## II. SELF-CONSISTENT MODEL OF DIRAC FERMION MOTION IN ELECTROMAGNETIC FIELD

### A. Dirac pseudo-spin and spatial dispersion of surface conductivity in CNT

In this section we will consider the self-consistent mechanism of CNT conductivity based on the model of Dirac pseudospins [73]. This theory covers the wide frequency range (from THz until visible light inclusively) and takes into account the interband transitions. The appropriate Hamiltonian for this case may be written as

$$\hat{H} = -\hbar v_F \hat{\sigma}_z \frac{\partial}{\partial z} - eE_z(t)\hat{z} \tag{1}$$

where the first term describes the free motion of Dirac fermions and the second one corresponds to their interaction with the EM-field. We will next solve the following Liouville equation for $2\times 2$ pseudospin matrix

$$i\hbar \frac{\partial \hat{\rho}}{\partial t} = \left[\hat{H}_K, \hat{\rho}\right] + \left[\hat{V}(t), \hat{\rho}\right] \tag{2}$$

where $\hat{H}_K = -\hbar v_F \hat{\sigma}(\partial/\partial z)$ ($\hbar$ is the reduced Planck's constant, $v_F$ represents the Fermi velocity and $\hat{\sigma}_z = (0, -i; i, 0)$ is the longitudinal Pauli matrix), $\hat{V} = -eE_z\hat{z}$ ($e$ denotes the electron charge, $E_z$ is the longitudinal component of electric field and $\hat{z}$ is the operator of longitudinal coordinate). The second term in (2) is considered to be small due to the linearity of CNT interaction with the EM-field and will be accounted for by using perturbation technique. We will consider a monochromatic EM-field, which means letting $\hat{V}(t) = \text{Re}(\hat{V}e^{-i\omega t})$ and $\hat{\rho}(t) = \hat{\rho}_0 + \delta\hat{\rho}e^{-i\omega t}$ where $\delta\hat{\rho}$ is taken as a small correction (perturbation) to the ambient EM-field. In addition, $\rho_0 = F(\varepsilon) = (1 + \exp((\varepsilon - \mu)/(k_B T)))^{-1}$ denotes the equilibrium density destribution (Fermi distribution), $\varepsilon$ is the electron's energy, $\mu$ is the electrochemical potential, $k_B$ repesents the Boltzmann constant and $T$ is the temperature.

Given that the spatial dispersion is weak, the electromagnetic fields and the induced currents, can be expressed as a superposition of traveling waves in the form $of \exp[i(qz - \omega t)]$, which propagate in opposite directions. Far from the CNT terminations (ends), their interaction can be accurately modeled using the corresponding infinite-length CNT approximation. Reflections from the extreme ends of a CNT and accounting for nonlocal effects, are incorporated by applying the ABCs. Following the Kubo formalism, the perturbation in the elements of the density matrix, can be represented in the momentum space as

$$(\delta\hat{\rho})_{mn,p,p+q} = \frac{F(\varepsilon_{m,p+q}) - F(\varepsilon_{np})}{\left[\varepsilon_{np} - \varepsilon_{m,p+q} + \hbar\omega + i0\right]} \hat{V}_{mn,p,p+q} \qquad (3)$$

where $\varepsilon_{m,p}$ is the energy of the electron with momentum $p$ corresponding to a zone with index $m$. The conductivity in the momentum space reads

$$\sigma_{zz}(\omega,q) = \frac{i}{\omega}\left[\Pi(\omega,q) - \Pi(0,q)\right] \qquad (4)$$

where

$$\Pi(\omega,q) = \sum_{mn}\sum_{p} \frac{F(\varepsilon_{m,p+q}) - F(\varepsilon_{np})}{\left[\varepsilon_{np} - \varepsilon_{m,p+q} - \hbar(\omega + i0)\right]} \left|\langle\hat{j}_z\rangle_{mn,p,p+q}\right|^2 \qquad (5)$$

is defined as the polarizability and $\langle\hat{j}_z\rangle_{mn,p,p+q}$ represent the matrix elements of the operators of the current density. The integerial sum over $p$ is a superposition over the states with different momenta and is conventional form of integration over the BZ zone. It is also convenient to split the total nonlocal conductivity into inter and intra components, namely, $\sigma_{zz}(\omega,q) = \sigma_{zz}^{inter}(\omega,q) + \sigma_{zz}^{intra}(\omega,q)$. The first term $\sigma_{zz}^{inter}(\omega,q) = i\omega^{-1}\Pi(\omega,q)$ in (4), corresponds to interband transitions whereas the second term $\sigma_{zz}^{intra}(\omega,q) = -i\omega^{-1}\Pi(0,q)$, corresponds to intraband transitions (also known as the Drude term in conductivity).

To facilitate the incorporation of the nonlocal effects, we employ a Taylor series expansion of (4) with respect to the momentum $q$ by retaining the leading correction term (that is, the second-order term in the series). Following [53], we will adopt the approximation $\langle \hat{j}_z \rangle_{mn,p,p+q} \approx \langle \hat{j}_z \rangle_{mn,pp}$ along with the following dispersion relation for Dirac fermions $\varepsilon_{mp} \approx \pm v_F p$. Since the current in the CNT is supported by contributions from all available states, (5) can also be reformulated as:

$$\Pi(\omega,q) = \sum_{mn}\sum_{p} F(\varepsilon_{np}) \left|\langle \hat{j}_z \rangle_{mn,pp}\right|^2 \left\{ \frac{1}{\left[\varepsilon_{np} - \varepsilon_{m,p+q} - \hbar(\omega+i0)\right]} + \frac{1}{\left[\varepsilon_{np} - \varepsilon_{m,p+q} + \hbar(\omega+i0)\right]} \right\} \quad (6)$$

which implies that

$$\sigma_{zz}^{\text{inter}}(\omega,q) = i\omega^{-1}\Pi(\omega,q) \approx i\omega^{-1}\left[\Pi(\omega,0) + \frac{1}{2}\left.\frac{\partial^2 \Pi(\omega,q)}{\partial q^2}\right|_{q=0} \cdot q^2\right] \quad (7)$$

Returning to the position space, we perform an inverse transformation $q \to \partial/\partial z$, which suggests that the total current can be expressed as the sum of local and nonlocal components. i.e.,

$$j_z = \sigma_{zz}(\omega,0) E_z + j_{z,\text{Nonloc}} \quad (8)$$

where

$$j_{z,\text{Nonloc}} = -\xi(\omega)\frac{\partial^2 E_z}{\partial z^2} \quad (9)$$

with the factor of nonlocality defined as

$$\xi(\omega) = \frac{1}{2}\left.\frac{\partial^2 \sigma_{zz}(\omega,q)}{\partial q^2}\right|_{q=0} \quad (10)$$

where for reasons of brevity shortness, we define $\sigma_{zz}(\omega,0) = \sigma_{zz}(\omega)$. The nonlocal component of the current may also be separated in a similar way into interband and intraband terms, thus the factor of nonlocality becomes $\xi(\omega) = \xi^{\text{inter}}(\omega) + \xi^{\text{intra}}(\omega)$. Combining (6), (7) and (10) finally renders

$$\xi(\omega) = 4i\hbar(\hbar v_F)^2 \sum_{mn}\sum_{p} \frac{\left(F(\varepsilon_{np}) - F(\varepsilon_{mp})\right)\left|\langle \hat{j}_z \rangle_{mn,pp}\right|^2}{\left[\varepsilon_{np} - \varepsilon_{mp} + \hbar(\omega + i0)\right]^3 (\varepsilon_{np} - \varepsilon_{mp})} . \tag{11}$$

Nonlocality is evident in (9) through the presence of the second-order spatial derivative, indicating that the current density at a given point depends not only on the value of the electric field at that point, but also on its spatial vicinity. Note that the term corresponding to the first-order derivative in the Taylor series vanishes due to axial symmetry of achiral CNTs, which ensures invariance under rotation around the tube axis. The value of the nonlocality parameter given in (11), relates to the class of general susceptibilities, while the value $\partial^2 E_z / \partial z^2$ is associated with the generalized force. The term $\xi(\omega)$ satisfies all general properties of general susceptibilities, as well as the analytical properties at the frequency in the complex plane [49]. Taking into acount that $\partial^2(x^{-1})/\partial x^2 = 2x^{-3}$, we obtain

$$\frac{2\hbar^2}{\left[\varepsilon_{np} - \varepsilon_{m,p+q} \pm \hbar\omega\right]^3} = \frac{\partial^2}{\partial \omega^2}\left(\frac{1}{\varepsilon_{np} - \varepsilon_{m,p+q} \pm \hbar\omega}\right) \tag{12}$$

and by virtue of (12), we can rewrite (11) in the form:

$$\xi(\omega) = -2i\hbar v_F^2 \frac{\partial^2}{\partial \omega^2} \sum_{mn}\sum_{p} \frac{\left(F(\varepsilon_{np}) - F(\varepsilon_{mp})\right)\left|\langle \hat{j}_z \rangle_{mn,pp}\right|^2}{\left[\varepsilon_{np} - \varepsilon_{mp} + \hbar(\omega + i0)\right](\varepsilon_{np} - \varepsilon_{mp})} \tag{13}$$

which immediately yields

$$\xi(\omega) = \frac{1}{2} v_F^2 \frac{\partial^2 \sigma_{zz}(\omega)}{\partial \omega^2} . \tag{14}$$

Thus, it is rather remarkable that the factor of nonlocality (spatial dispersion) in our model is uniquely defined by the second-order derivative of the local conductivity with respect to the frequency (temporal dispersion). This relation constitutes the main result of this Section, which will be used in a sequel as a framework for our analysis. The symbol $\omega + i0$ is conventionally associated with attenuation. In order to calculate real part of the polarizability (6) (and conductivity), one must perform the following Sokhotski-Plemelj formula [73]:

$$\text{Im}\left(\frac{1}{\varepsilon_n - \varepsilon_m - \hbar\omega - i0}\right) \Rightarrow -\pi\delta(\varepsilon_n - \varepsilon_m - \hbar\omega) \tag{15}$$

which, in view of (6) finally renders

$$\text{Re}[\sigma_{zz}(\omega)] = \frac{\pi}{\omega}\sum_{mn}\sum_{p}(F(\varepsilon_{mp}) - F(\varepsilon_{np}))\delta(\varepsilon_{np} - \varepsilon_{mp} - \hbar\omega)|\langle \hat{j}_z \rangle_{mn,pp}|^2 \tag{16}$$

Note that the spatial dispersion term contributes to the collisionless relaxation in a CNT. By virtue of (14), it is evident that $\xi(\omega)$ is indeed a complex quantity and its real part can be easily found from (14) and (13). It is also worth noting in conjuction with (14), that the second-order derivative of a delta function corresponds to a distribution that maps an arbitrary test function to the second derivative of that function evaluated at a given point, i.e., $\int f(x)\delta''(x-a)dx = f''(a)$ [75].

### B. Optical far-infrared properties of CNT

Let us simplify the general Kubo formalism for low-energy excitations constrained by the condition $E < 1\,\text{eV}$. For this purpose, we adapt it to the case of a pseudo-spin electron liquid, similarly to the approach used for a monolayer graphene in [37]. To this end, we consider the electronic states near the K-points and employ a linear approximation for $\varepsilon_\mathbf{p} = \pm v_F|\mathbf{p} - \mathbf{p}_F|$. We will also consider the case of low temperature ($T \ll \mu$) and collision-free electron motion ($\nu = 0$). In contrast with graphene [37], [73], for CNT we take into acount the transverse electron confinement (the azimuthal component of electron momentum is discrete, while the longitudinal component is continuous).

Following [56], we refine the Kubo formalism to suit the specific model under consideration. For a zigzag CNT (*m*,0), we have

$$\sigma_{zz}^{\text{inter}}(\omega) = \frac{ie^2\omega}{2\pi^2\hbar R_{CN}}\sum_{s=1}^{m}\int_{BZ}\frac{|v_{cv}(p_z,s)|^2}{\varepsilon(p_z,s)} \cdot \frac{F[-\varepsilon(p_z,s)] - F[\varepsilon(p_z,s)]}{\hbar^2(\omega+i0)^2 - 4\varepsilon^2(p_z,s)}dp_z \tag{17}$$

where $v_{cv}(p_z,s)$ denotes a velocity matrix element of direct interband transitions [56]. For future consideration, it may be simply approximated by $v_{cv}(p_z,s) \approx \hbar v_F$ [56].

We will consider two types of estimations: The first one is the case of a rather small *m* (*m*<60). The main contribution to the conductivity in this case is provided by the terms *s*=*m*/3 and *s*=2*m*/3, which include Fermi points correspond to the different valleys of the BZ zone. Next, we transform the integration over $p_z$ to that over energy and carry it out over the range $0 < \varepsilon < \infty$, which renders

$$\sigma_{zz}^{\text{inter}}(\omega) = \frac{ie^2 \hbar \omega v_F}{\pi^2 R_{CN}} \int_0^\infty \frac{F(-\varepsilon) - F(\varepsilon)}{\left[\hbar^2(\omega + i0)^2 - 4\varepsilon^2\right]^2} \frac{d\varepsilon}{\varepsilon} \tag{18}$$

The principal contribution to the integral in Eq. (18) originates from the vicinity to the pole $\varepsilon \approx \hbar \omega$. The energy in the denominator may be approximated by this relation and removed out of the integral leading to the following approximated expression for (18):

$$\sigma_{zz}^{\text{inter}}(\omega) = \frac{ie^2 v_F}{\pi^2 R_{CN}} \int_0^\infty \frac{F(-\varepsilon) - F(\varepsilon)}{\left[\hbar^2(\omega + i0)^2 - 4\varepsilon^2\right]^2} d\varepsilon \tag{19}$$

The integral in (19) can be calculated using the same method as outlined in [39]. For zero temperature we have

$$\sigma_{zz}^{\text{inter}}(\omega) = H(\hbar\omega - 2\mu) + i\frac{\sigma_0}{\pi}\left(\frac{4\mu}{\hbar\omega} - \frac{1}{2}\ln\left(\frac{\hbar\omega + 2\mu}{\hbar\omega - 2\mu}\right)^2\right) \tag{20}$$

where $H(x)$ represents the Heaviside step function and $\sigma_0 = e^2/4\hbar$ is the quantum of resistivity. For the parameter of nonlocality $\xi$ we obtain folowing (14) and (19)

$$\xi^{\text{inter}}(\omega) = i\frac{16\sigma_0 \mu v_F^2}{\pi \hbar \omega^3}\left\{1 - \frac{(\hbar\omega)^4}{\left[(\hbar\omega)^2 - 4\mu^2\right]^2}\right\} \tag{21}$$

The singularity in (20) and (21) at the step point $\hbar\omega - 2\mu$, will be regularized by introducing finite temperature effects.

The second estimation corresponds to high values of $m > 200$ (exactly, the limit of graphene). In this case the integration procedure is as follows: The limit $m \to \infty$ means the exchange of the sum in (17) by an integral with respect to $\hbar(2\pi R_{CN})^{-1}\sum_{s=1}^m (...) \to \int_0^{2\pi}(...)dp_\phi$, where $p_\phi$ is the azimuthal component of the momentum. The integration performed in the momentum space in the continuous limit, is transformed to integration over the energy $\varepsilon$ and $\phi$ by extending the integration domain over $0 < \varepsilon < \infty$. The differential with respect to the momentum, can be then replaced in (17) by $d^2\mathbf{p} = v_F^{-2}\varepsilon d\varepsilon d\phi$, leading to

$$\sigma_{zz}^{inter}(\omega) = \frac{ie^2\omega}{\pi}\int_0^\infty \frac{F(-\varepsilon) - F(\varepsilon)}{\left[\hbar^2(\omega+i0)^2 - 4\varepsilon^2\right]^2} d\varepsilon \qquad (22)$$

The above relation is in exact agreement with the corresponding expression for the graphene conductivity derived in [37], [39]. The correspondence between conductivities of CNTs and graphene, is clearly established through the following exchange $v_F/\pi R_{CN} \leftrightarrow \omega$.

### III. EIGENMODES IN CNT WITH SPATIAL DISPERSION

For a deeper physical understanding of the results that follow, it is instructive to examine the eigenmodes of an infinitely long CNT exhibiting spatial dispersion, represented in the form of a traveling- wave, with $h$ as the guide wavenumber to be determined. The analysis is confined to azimuthally symmetric modes, as azimuthally dependent modes, are of lesser relevance within the scope of the current investigation. The electromagnetic field is characterized by the longitudinal component of the electric Hertz vector $\Pi$

$$E_z = (k^2 - h^2)\Pi \qquad (23)$$

$$H_\varphi = i\varepsilon_0 \omega \frac{\partial \Pi}{\partial r}. \qquad (24)$$

Using (6) and (11), one gets

$$\frac{\partial^2 j_z}{\partial z^2} + \tilde{\alpha}^2(\omega) j_z = \frac{\sigma_{zz}^2(\omega)}{\xi(\omega)}(k^2 - h^2)\Pi \qquad (25)$$

where $\tilde{\alpha}(\omega) = \sqrt{\sigma_{zz}(\omega)/\xi(\omega)}$.

The current is excited by the following longitudinal component of the electric field on the CNT surface. Note that the electric field generated by the surface current satisfies (25). This field is governed by the three-dimensional wave equation derived from Maxwell's equations, which can be expressed in terms of the electric Hertz vector as

$$(\nabla^2 + k^2)\Pi = -\frac{i}{\omega\varepsilon_0}\delta(r - R_{CN}) j_z \qquad (26)$$

where $\delta(x)$ is the Dirac function.

We can now proceed to formulate the effective boundary conditions prevailing on the CNT surface. The discontinuity in the tangential component of the magnetic field is proportional to the surface current density, while the tangential component of the electric field remains continuous across the surface. However, in contrast to the conventional case of local conductivity, these boundary conditions are modified due to the presence of the leading nonlocal term in (25), which reflects the spatially dispersive nature of the response. Denoting the interior ($R_{CN} - 0$) and the exterior ($R_{CN} + 0$) components of the Hertz vector by $\Pi^{(\pm)} = \Pi_{r=R_{CN}\pm 0}$, the effective boundary conditions for the eigenmodes are

$$\Pi^{(+)} - \Pi^{(-)} = 0, \tag{27}$$

$$\left(1 - \frac{h^2}{\tilde{\alpha}^2}\right)\frac{\partial}{\partial r}\left(\Pi^{(+)} - \Pi^{(-)}\right) = \frac{\sigma_{zz}(\omega)}{i\omega\varepsilon_0}\left(k^2 - h^2\right)\Pi, \tag{28}$$

(The subscript $\pm$ in the right-hand part of (28), is omitted because of continuity). The solution of the homogeneous Helmholtz solution for the Hertz potential, can be expressed in terms of the modified Bessel functions as

$$\Pi(r,z) = \begin{cases} I_0(\nu r)K_0(\nu R_{CN}), r < R_{CN} \\ K_0(\nu r)I_0(\nu R_{CN}), r > R_{CN} \end{cases} e^{ihz} \tag{29}$$

where $h = \pm\sqrt{\nu^2 + k^2}$. The solution of (29) satisfies the boundary condition (27) for the continuity of the tangential electric field component at the CNT surface. Applying the boundary condition (28), renders the following characteristic equation:

$$\left(\frac{\nu}{k}\right)^2 I_0(\nu R_{CN}) K_0(\nu R_{CN}) = \frac{i}{kR_{CN}Z_0\sigma_{zz}(\omega)}\left(1 - \frac{k^2 + \nu^2}{\tilde{\alpha}^2}\right) \tag{30}$$

where $Z_0$ denotes the wave impedance of vacuum.

Next, we will consider this equation with respect to a transverse wavenumber $\nu$ for a given $k$. For given values of the complex conductivity $\sigma_{zz}(\omega)$ and $\xi(\omega)$, the roots of (30) are complex valued. Solutions of (30) that reflect a valid physical meaning, must satisfy the condition $\text{Re}(\nu) > 0$ (i.e., the field dissipates over the propagation along CNT). To simplify the analysis, we will assume large values of the wave-number X (i.e., 'non-slender' CNT), so that to leading order $2I_0(X)K_0(X) \approx X^{-1}$. In this case, (30) yields two roots

$$v_{1,2}(\omega) = \frac{\tilde{\alpha}^2 Z_0 \sigma_{zz}(\omega)}{4ik}\left(-1 \pm \sqrt{1+\left(\frac{4k}{\tilde{\alpha} Z_0 \sigma_{zz}(\omega)}\right)^2\left(\frac{k^2}{\tilde{\alpha}^2}-1\right)}\right) \qquad (31)$$

Each root of (31) corresponds to the pair of waves travelling in opposite directions. The transition to the case where spatial dispersion is absent is defined by the asymptotic behavior of (31) in the limit of $|\tilde{\alpha}| \to \infty$, which results in

$$v_1 \to \frac{2ik}{Z_0 \sigma_{zz}(\omega)} \qquad (32)$$

$$v_2 \to -\frac{\tilde{\alpha}^2 Z_0 \sigma_{zz}(\omega)}{2ik}. \qquad (33)$$

It is important to note that the first root (32) aproaches a finite value that corresponds to the ordinary plasmon-polaritons, which are well-known in CNTs [56]. Such modes exist only in the frequency range for which $\text{Im}(\sigma_{zz}(\omega)) > 0$. The second root (33), tends to infinity, which means an increase of mode confinement in the vicinity of the CNT surface. Therefore, the energy flux approaches zero for the constant amplitude value, which means that these modes don't exist in the limit of local conductivity and should be ignored in this case. It is also worth mentioning that modes corresponding to large values of the wave-number are unphysical, since their existence contradicts the prior assumption of weak spatial dispersion. However, the roots (33) create the real physical modes for the small values of $\text{Im}(\sigma_{zz}(\omega))$ and exist for $\text{Im}(\sigma_{zz}(\omega)) < 0$ (where the roots (32) and (33) are comparable). Such modes exist for $\text{Im}(\sigma_{zz}(\omega)) < 0$ and hold at the immediate vicinity of the conductivity resonances. Such solutions correspond to the additional modes which are produced by the spatial dispersion. These solutions, that are manifested by the new types of resonances in the scattering process and observed in the numerical simulations, must be taken into account in the present methodology.

The situation is rather similar to that in crystal optics [24], where the spatial dispersion creates the additional modes and the permittivity tensor is given by $\varepsilon_{\alpha\beta}(\mathbf{k};\omega) = \varepsilon_{\alpha\beta}^{(0)}(\omega) + \gamma_{\alpha\beta\chi\delta}(\omega)k_\chi k_\delta$ ( $\gamma_{\alpha\beta\chi\delta}(\omega)$ is given factor of non-locality, $k_{\chi,\delta}$ are the Cartesian components of the wavevector $\mathbf{k}$;

the sums over doubly repeated indexes is assumed) [24]. The inclusion of wavevector-dependent terms in the expression for the permittivity, elevates the order of the algebraic dispersion relation, thereby introducing additional roots corresponding to the new wave modes. Far from the absorption line, these modes exist in the range of large **k,** i.e., outside the area of validity of the theory and thus must be discarded. Near the absorption line, the permittivity varies considerably even for small values of **k.** For this reason, these newly found additional roots are of real physical significance and must be included in the formulation.

## IV. ADDITIONAL BOUNDARY CONDITIONS IN CNT
### A. Charges, currents, fields

To fully specify the problem under consideration, it is essential to characterize the behavior of some of the key physical quantities—namely, current and charge densities as well as the components of the electromagnetic field in the vicinity of the extreme ends of the CNT. This section addresses this issue by employing a model of a semi-infinite CNT and applying the WH technique [51]– [55]. The WH method is particularly suited for convolution-type integral equations that are defined over semi-infinite domains. In some contexts, it is applied directly to the Helmholtz equation, in which case it is often referred to as the Jones method. Despite some differences in the formulation, both approaches are mathematically equivalent. Here, we employ the exact analytical solution derived for the wave-scattering problem by a semi-infinite CNT, mainly to examine the field behavior near the nanotube edge. The geometry of the considered problem is shown at Fig 1. It is convenient to separate the total EM field into an incident (inc) and scattered (sc) components. The physically relevant regime corresponds to the case where the radius of the CNT can be taken as small compared to the wavelength of the incident electromagnetic field ($kR_{CN} \ll 1$). Hence, the variation of the longitudinal component of the incident field across the CNT cross-section maybe neglected, allowing its value to be approximated by its corresponding value on the CNT axis. The transverse component, when averaged over the CNT surface, does not contribute significantly to the induced current. Consequently, the induced current density can be treated as azimuthally symmetric, and the resulting scattered field will also exhibit azimuthal independence. In the sequel we consider the scattering field in the form of a plane wave $E_z^{inc} = E_0 \exp(i(kz\cos\theta_0 - \omega t))$, $E_0 = -\tilde{E}_0 \sin\theta_0$, where $\tilde{E}_0$ denotes the constant amplitude and $\theta_0$ is the incident angle.

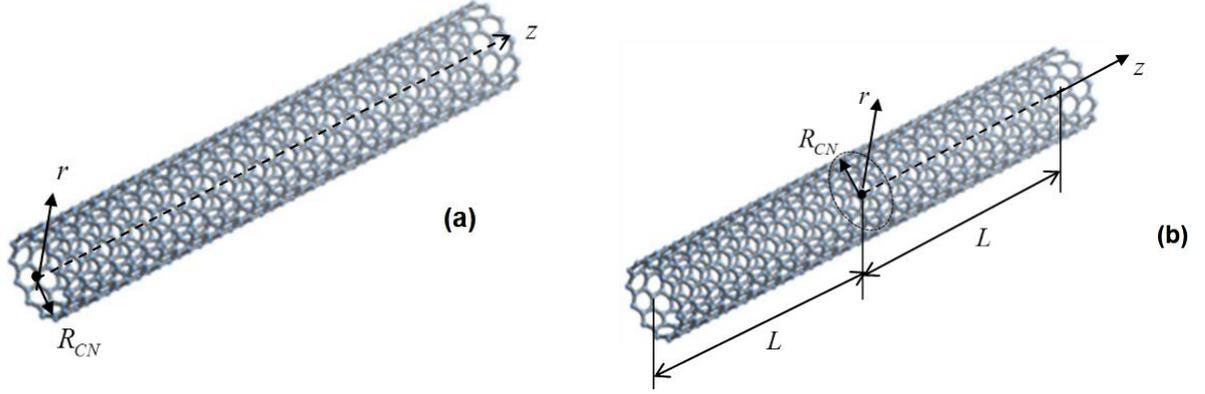

FIG. 1. Geometry of the problems under consideration with their corresponding cylindrical origins indicated; **(a)** semi-infinite CNT. **(b)** finite-length CNT.

The field scattered by the CNT, is described by the longitudinal component of the electric Hertz vector $\Pi$, in a similar manner to the eigenmode derivation outlined in Section III (see also (27) and (28)), resulting in

$$E_z^{sc} = \frac{\partial^2 \Pi}{\partial z^2} + k^2 \Pi \ . \tag{34}$$

$$H_\varphi^{sc} = i\varepsilon_0 \omega \frac{\partial \Pi}{\partial r} \tag{35}$$

By enforcing (5) and (7), one gets

$$\frac{\partial^2 j_z}{\partial z^2} + \tilde{\alpha}^2(\omega) j_z = \frac{\sigma_{zz}^2(\omega)}{\xi(\omega)} \left( \frac{\partial^2 \Pi}{\partial z^2} + k^2 \Pi + E_z^{inc} \right) \tag{36}$$

where $\tilde{\alpha}(\omega) = \sqrt{\sigma_{zz}(\omega)/\xi(\omega)}$. The Hertz vector in (36) satisfies (26), which is equivalent to using Maxwell's equations for the scattered field. The corresponding effective boundary conditions constitute a mixed-boundary problem, as their form differs between the CNT surface and the surrounding medium, i.e.,

$$\Pi^{(+)} - \Pi^{(-)} = 0, -\infty < z < \infty \tag{37}$$

$$\left(1+\frac{1}{\tilde{\alpha}^2}\frac{\partial^2}{\partial z^2}\right)\frac{\partial}{\partial r}\left(\Pi^{(+)}-\Pi^{(-)}\right)=\frac{\sigma_{zz}(\omega)}{i\omega\varepsilon_0}\left(\frac{\partial^2}{\partial z^2}+k^2\right)\Pi+\frac{\sigma_{zz}(\omega)}{i\omega\varepsilon_0}E_z^{inc}, 0<z<\infty \quad (38)$$

and

$$\frac{\partial}{\partial r}\left(\Pi^{(+)}-\Pi^{(-)}\right)=0, -\infty<z<0 \quad (39)$$

(The subscript $\pm$ in the right-hand part of (38) is again omitted due to continuity). The problem is thus formulated as follows: solve Maxwell's equations for the scattered field subject to the boundary conditions (37-38) on the CNT surface and (38-39) outside the surface, including the radiation condition at infinity and the Meixner's concept of edge condition [76]. The Meixner condition asserts that the electromagnetic energy stored within any finite region of space must always remain bounded, thereby excluding the appearance of unphysical point sources or singularities at the CNT ends.

The detailed deduction of the WH equation is given in Appendix A and the solution of the problem is finally presented in the form of a Fourier integral

$$\Pi(r,z)=\frac{1}{2\pi}\int_{-\infty}^{\infty}\bar{\Pi}(r,\alpha)e^{-i\alpha z}d\alpha \quad (40)$$

where $\bar{\Pi}(r,\alpha)$ represents the spectral amplitude to be found. Following the Helmholtz equation and the radiation condition we obtain

$$\bar{\Pi}(r,\alpha)=\begin{cases}A(\alpha)K_0(\gamma r), r>R_{CN}\\ D(\alpha)I_0(\gamma r), r<R_{CN}\end{cases} \quad (41)$$

where $\gamma=\sqrt{\alpha^2-k^2}$, $A(\alpha), D(\alpha)$ are some unknown functions and $I_0(\gamma r), K_0(\gamma r)$ are the modified Bessel and Macdonald's functions, respectively.

Let us now perform an analytic transformation into the complex plane of the variable $\alpha$. Since multivalued functions are involved, the complex variable $\alpha$ is defined at the two-sheeted Riemann surface characterized by the branch points $\pm k$, as depicted at Fig. (2a). The function $\bar{\Pi}(r,\alpha)$ may be decomposed into the form $\bar{\Pi}(r,\alpha)=\bar{\Pi}_+(r,\alpha)+\bar{\Pi}_-(r,\alpha)$, where $\bar{\Pi}_+(r,\alpha)=\int_0^{\infty}\Pi(r,z)e^{i\alpha z}dz$ and $\bar{\Pi}_-(r,\alpha)=\int_{-\infty}^0\Pi(r,z)e^{i\alpha z}dz$. As one can see, the functions $\bar{\Pi}_\pm(r,\alpha)$ define analytical functions in top and bottom half-planes of $\alpha$. It is also important to note that the Fourier transformation of the current density contains only the analytic component at the top half-plane, since the current exists only for positive z values.

Finally, recalling that the boundary conditions (37)-(39) lead to the following WH equation

$$\Upsilon_+(\alpha)G(\alpha) = \Xi_-(\alpha) + \frac{1}{i\pi}\breve{\Phi}(\alpha) \tag{42}$$

where $\Upsilon_+(\alpha), \Xi_-(\alpha)$ are unknown functions of a complex variable $\alpha$ which are analytic in top and bottom half-planes, respectively ($\Upsilon_+(\alpha)$ is the Fourier transformation of the current density) and the functions $G(\alpha), \breve{\Phi}(\alpha)$ are prescribed, we get

$$G(\alpha) = I_0(\gamma R_{CN})K_0(\gamma R_{CN}) + \frac{1}{\zeta\gamma^2}\left(1 - \frac{\alpha^2}{\tilde{\alpha}^2}\right) \tag{43}$$

where $\zeta = i\pi R_{CN}\sigma_{zz}Z_0$, $Z_0$ is the free space impedance and the function $\breve{\Phi}(\alpha)$ is the Fourier transformation of the incident field. Equation (42), known as the WH equation, is expressed in terms of the two unknown functions. In order to ensure a unique solution, additional information is required in the form of specified regions of analyticity for each function. The WH equation admits an exact analytical solution (see Appendix A), which enables the computation of the relevant physical parameters, such as the induced currents, charge distributions, and components of the EM field.

The analytical solution of (42) is found based on the so-called factorization of the function $G(\alpha)$, i.e., its presentation in the form of a product $G(\alpha) = G_+(\alpha) \cdot G_-(\alpha)$, where $G_\pm(\alpha)$ are the analytical functions at the top and bottom half-planes, respectively (for further details see Appendix A). The final results for the Hertz potential (A18), electric field components (A19), (A20), are detailed in Appendix A, current density and charge density are summarized below:

$$j_z(z) = \frac{\omega\varepsilon_0 \eta}{kR_{CN}} \int_C \frac{e^{-i\alpha z}}{(\alpha + k\cos\theta_0)(\alpha + k)G_+(\alpha)} d\alpha \tag{44}$$

$$\rho(z) = -\frac{\eta\varepsilon_0}{kR_{CN}} \int_C \frac{\alpha}{(\alpha + k)(\alpha + k\cos\theta_0)G_+(\alpha)} \frac{e^{-i\alpha z}}{} d\alpha \tag{45}$$

where $\eta = \dfrac{E_0}{2\pi(1+\cos\theta_0)G_+(k\cos\theta_0)}$ is an amplitude factor and $C$ represents the integration contour shown in Fig. 2b. Note that equation (45) for the charge density may also be obtained from (44) by virtue of the continuity relation.

## B. Physical analysis

The exact solution of the problem presented in the previous section takes the form of a standard Fourier integral $F(z) = \int_C S(\alpha) e^{-i\alpha z} d\alpha$. It is well-known that it's asymptotic behavior at $z \to 0$, is dictated by the asymptotic behavior of its spectral amplitude $S(\alpha)$ at $|\alpha| \to \infty$ [42]. For the subsequent analysis, it is more convenient to recast these integrals into an alternative form. Such a transformation must account for the analytic properties of the integrands, each of which possesses three simple poles, i.e., $\alpha_j = \sqrt{v_j^2 + k^2}$ ($j=1,2$), corresponding to the roots of the characteristic equation (31)) and $\alpha_3 = -k\cos\theta_0$.

As an example, let us focus on (45), which describes the charge density. By shifting the strip from the branch point $-k$ until $-k - i\infty$, the contour of integration will be transformed to $P$ along this sheet, as shown in Fig. 2b. In this case, we cross the singular points and their residues in principle should be also accounted. These singular points correspond to the far-field and do not affect the field near the edge. Introducing next the new variable $u = i(k + \alpha)$, Eq. (45) can be tranformed to

$$\rho(z) = -2\frac{\eta\varepsilon_0}{kR_{CN}} e^{-ikz} \int_0^\infty \frac{(k+iu)G_-(k-iu)e^{-uz}}{[u - ik(1-\cos\theta_0)]G(k-iu)} du. \qquad (46)$$

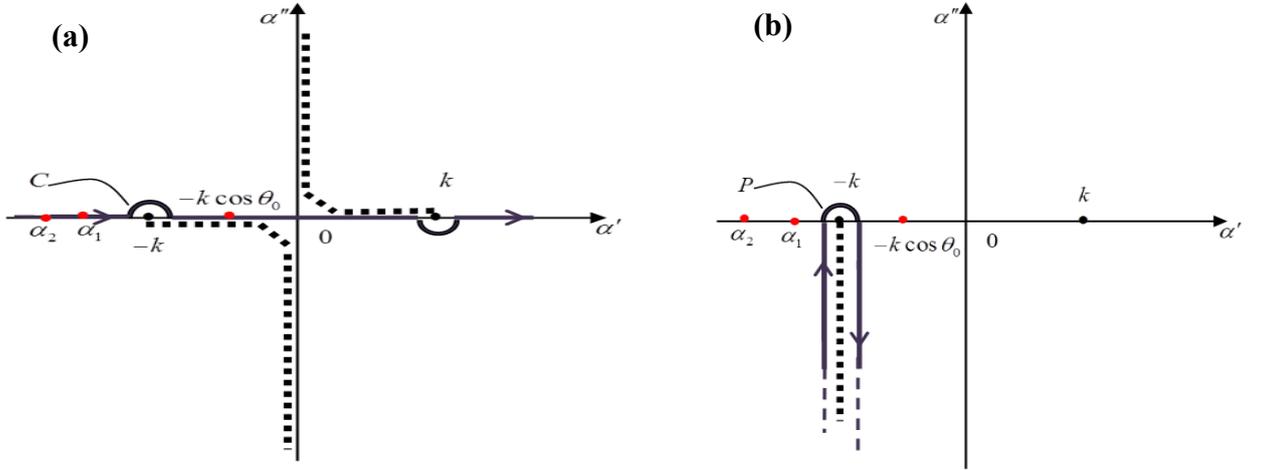

FIG. 2. **(a)** The complex plane $\alpha$ defined at the two-sheeted Riemann surface with branch points $\pm k$. The sheets are shown by dashed lines. Three simple poles of integrating functions (colored red) are given by $\alpha_j = \sqrt{v_j^2 + k^2}$ ($j=1,2$), given in terms of the roots of (30) and $\alpha_3 = -k\cos\theta_0$. The contour of integration is denoted by $C$. **(b)** The strip shifted from the branch point $-k$ until $-k - i\infty$. The contour of integration circumventing $P$ along this sheet.

The main outcome of the obtained solution, is the modification of the asymptotic behavior of the spectral densities in the integrals (A18)–(A20), (44), and (45), as a consequence of spatial dispersion. Indeed, in the absence of nonlocal effects, Eq. (52) should be evaluated by taking the limit $\tilde{\alpha} \to \infty$ for the nonlocality parameter. The agreement with the Meixner condition [73] (finite energy stored in the finite space), allows the singular behavior $O\left(\left(\rho^2+z^2\right)^{-\phi}\right)$ of some field components with $\phi \leq 1/2$. In this case, the charge density behaves like $\rho(z) = O(z^{-\phi})$. According to the asymptotic relation $I_0(\gamma R_{CN}) K_0(\gamma R_{CN}) \cong 1/2\gamma$ for $\gamma \to \infty$, the main contribution to Eq. (43), is due to its first term. As a result, we find that $G(\alpha) = O(\alpha^{-1})$ and $G_+(\alpha) = O(\alpha^{-1/2})$. Therefore, we set $\phi = 1/2$ in (46), implying that the charge density is $\rho(z) = O(1/\sqrt{z})$. Using similar arguments, suggest that the current density $S(\alpha) = O(|\alpha|^{-3/2})$ and $j_z(z) = O(\sqrt{z})$. Such an asymptotic behavior agrees with the continuity relation. The electric field components in (A19) and (A20) exhibit the same order of singularity. The situation changes remarkably when the spatial dispersion (i.e., nonlocality) is taken into consideration. In this case, the quantity remains finite, and the dominant contribution to Eq. (43) arises from its last term. As a result, the current density is $S(\alpha) = O(|\alpha|^{-2})$ and

$$j_z(z) = O(z) \tag{47}$$

Thus, the singularities in both charge density and electric field may be ignored. However, the singular behavior of the eigenvalues in the local case emerges due to their unbounded growth near the ends (though only finite values have physical meaning).

Accurate physical modeling of the current density is crucial from a computational perspective. As previously discussed, the conventional formulation of the Pocklington integral equation for thin wires [57], [58] involves an interchange between integral and differential operators. However, these operators are inherently non-commutative. This mathematical simplification renders the resulting equations ill-posed and ultimately unsolvable, giving rise to the spurious oscillations that arise in the computed current distribution, especially near the wire ends. Such numerical artifacts stem from neglecting the correct physical behavior of the current at the vicinity of the ends [60], [61]. A similar scenario also arises in the classical analysis of the EM field near an impedance half-plane, a problem which has been extensively studied in the context of edge effects [51]. This configuration is commonly used for example to model the skin effect in edge structures and to estimate power losses in various microwave devices. From a physical perspective, the impedance boundary conditions become invalid in the immediate vicinity of the edge, because they presuppose that the surface curvature is small compared to the skin depth. At the edge, however, the curvature becomes infinite, thus invalidating this assumption and as a result lead to errors in the modeled current distribution. Although the region of significant error is spatially confined, accurate estimation of integral losses still requires a correct modeling of the current behavior near the CNT edges.

Traditionally, the perturbation method is employed to estimate energy losses by approximating the fields with their counterparts from a perfectly conducting structure of identical geometry [51]. This idealization, which assumes zero surface impedance, inherently introduces singularities into the field distribution, leading to some divergent integrals when calculating the dissipated energy. For practical numerical implementations, these singularities can be artificially replaced by corresponding finite values that arise from the computational artifacts rather than any underlying physical meaning. As a result, the numerical evaluation of energy losses in such cases becomes ill-defined and may yield some unphysical results. In contrast, when the concept of a finite surface impedance is properly incorporated in the analysis, as is naturally done in nonlocal electrodynamic formulations, these singularities are inherently suppressed. The corresponding integrals for energy dissipation thus become convergent, yielding stable and physically consistent estimates.

## V. INTEGRAL EQUATION FOR THE CURRENT DENSITY IN A CNT WITH NONLOCAL CONDUCTIVITY

The current is excited by the longitudinal component of the electric field evaluated on the CNT surface, which satisfies Eq. (34). The system is governed by a 2D wave equation derived from Maxwell's equations that is expressed in terms of the electric Hertz vector as in Eq. (40). Together, Eqs. (36) and (26) formulate a self-consistent problem describing the interaction of a finite-length carbon nanotube (CNT) with an external electromagnetic (EM) field. This results in a coupled system of wave equations for both the electromagnetic field and the surface current. Specifically, the electromagnetic field throughout the surrounding space is generated by the current on the CNT surface, while the surface current itself is induced by the total electric field. By inverting the differential operator in the left-hand part of Eq. (36) with boundary conditions (37)-(39), we obtain the following nonlocal conductivity law:

$$j_z(z) = -\sigma_{zz}(\omega)\tilde{\alpha}^2(\omega)\int_{-L}^{L} g(z,z')E_z(z')dz' \qquad (48)$$

where

$$g(z,z') = -\frac{1}{\tilde{\alpha}\sin(2\tilde{\alpha}L)}[H(z'-z)\sin(\tilde{\alpha}(z+L))\sin(\tilde{\alpha}(z'-L)) + \qquad (49)$$
$$H(z-z')\sin(\tilde{\alpha}(z'+L))\sin(\tilde{\alpha}(z-L))]$$

is the corresponding 1D Green function for the 1D Helmholtz equation satisfying the boundary conditions given by Eq. (47), namely $j_z(-L) = j_z(L) = 0$, where $H(z)$ denotes the Heaviside step function. Some other equivalent forms of this Green function are also known, however we find that this one is more convenient for the present formulation. The relation $\tilde{\alpha}(\omega_n) = n\pi/2L$ defines the eigenfrequencies of the additional modes created by nonlocality. It is also instructive to note, that the kernel $g(z,z')$ does not have the common convolution form $g(z-z')$, due to the finite-length of the model under consideration. Let us next express the Hertz potential as

$$\Pi(r,z) = \frac{i}{\omega\varepsilon_0}\int_{-L}^{L} j_z(s) G(r,z-s) ds \qquad (50)$$

where $j_z(z)$ is the unknown current density and the kernel is given by

$$G(r,s-z) = R_{CN}\int_0^{2\pi}\frac{e^{ik\sqrt{r^2+R_{CN}^2-2rR_{CN}\sin(\phi)+(s-z)^2}}}{\sqrt{r^2+R_{CN}^2-2rR_{CN}\sin(\phi)+(s-z)^2}}d\phi . \qquad (51)$$

Substituting Eq. (48) into Eq. (50) yields the exact solution to Maxwell's equations throughout the entire space, satisfying both the radiation condition and the boundary condition expressed in Eq. (38). The surface current density appearing in Eq. (50) must then be determined using the boundary condition (39). By substituting Eq. (50) into Eq. (39), we arrive at the governing equation for the current density

$$j_z(z) + \frac{i\sigma_{zz}(\omega)\tilde{\alpha}^2(\omega)}{\omega\varepsilon_0}\int_{-L}^{L} g(z,z')\left(\frac{\partial^2}{\partial z'^2}+k^2\right)\left\{\int_{-L}^{L} j_z(s)G(s-z')ds\right\} dz' = B(z) \qquad (52)$$

where $G(s-z') = G(R_{CN}, s-z')$ is given by Eq. (51) and

$$B(z) = \sigma_{zz}(\omega)\tilde{\alpha}^2(\omega)\int_{-L}^{L} g(z,z') E_z^{inc}(z') dz' . \qquad (53)$$

The right-hand side of Eq. (53) is defined in terms of the incident field. From an operator-theoretic perspective, this expression can be regarded as an integro-differential equation involving the product of three operators: the first and third are integral operators with kernels $g(z,z')$ and $G(s-z')$, respectively, and the second one is the differential operator $\partial_{z'}^2 + k^2$ of the Sturm-Liouville type. In its present form this integral expression seems to be unsuitable for direct numerical treatment and thus must be reformulated into the canonical form of a Fredholm integral equation to facilitate stable and efficient computational implementation. One of the conventional approaches along these lines is based on changing the order of integration over $dz'$ and $ds$, noting that $\partial^2/\partial z'^2 = \partial^2/\partial s^2$ and making use of the following approximation:

$$G(s-z) \approx 2\pi R_{CN} \frac{e^{ik\sqrt{4R_{CN}^2+(s-z)^2}}}{\sqrt{4R_{CN}^2+(s-z)^2}} \tag{54}$$

Such an approach is widely used in the theory of macroscopic wire antennas and is often named the Pocklington equation [54]. However, as noted by Hanson et.al [63], the kernel $G(z'-s)$ is expressed in terms of a singular integral. Therefore, the change in the order of integration and differentiation is not appropriate (namely, the integral obtained in such a way is non - converging) and the resulting integral equation is ill-posed. Thus, its direct numerical solution using certain nonstandard methods, often results in some unphysical oscillations of the current density, particularly near the CNT boundaries. A renormalization approach aimed to mitigate such numerical artifacts was proposed by Hanson et al. [63]. Although their technique is not directly applicable to our case, due to the differences in the equation's structure, the underlying principle remains valuable. By dopting this approach, we derive the renormalized equation in the following form:

$$j_z(z) - \frac{i\sigma_{zz}(\omega)\tilde{\alpha}^2(\omega)}{\omega\varepsilon_0} \int_{-L}^{L} j_z(s) K(z,s) ds = B(z) \tag{55}$$

where

$$K(z,s) = \left(\frac{\partial^2}{\partial s^2} + k^2\right) \int_{-L}^{L} g(z,z') G(s-z') dz' = F_1(z,s) - F_2(z,s) + F_3(z,s) \tag{56}$$

$$F_1(z,s) = k^2 \text{V.P.} \int_{-L}^{L} g(z,z')G(s-z')dz' \tag{57}$$

$$F_2(z,s) = g(z,s)\frac{\partial}{\partial s}\left(G(s-L) - G(s+L)\right) \tag{58}$$

$$F_3(z,s) = \text{V.P.} \int_{-L}^{L} \left(g(z,z') - g(z,s)\right)\frac{\partial^2 G(s-z')}{\partial s^2}dz' \tag{59}$$

and the symbol V.P. denotes the Cauchy principal value of the integral. A detailed derivation of Eqs. (57)-(59) is otlined in Appendix B. It is also important to note that $F_2(z,\pm L) = 0$ and that the current density defined by Eq. (55), exactly satisfies the end conditions (47).

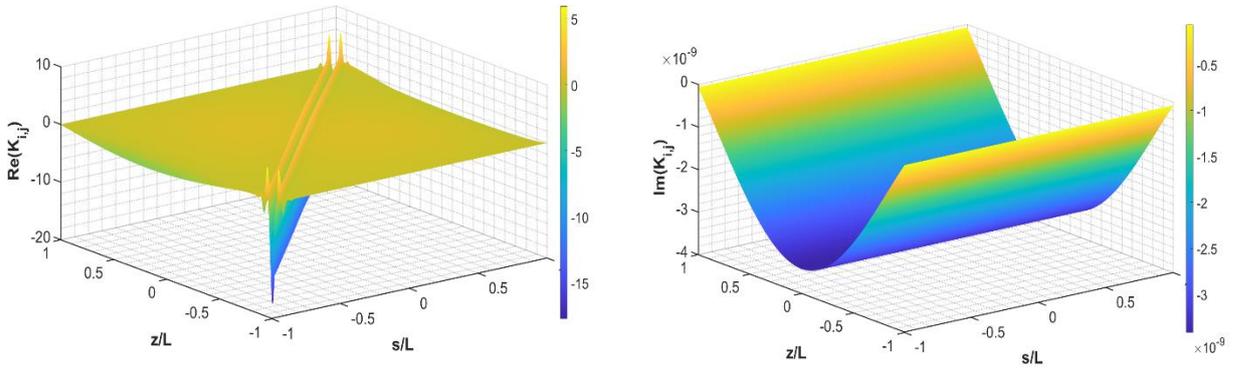

FIG. 3. The typical form of the kernel $K(z,s)$ of the integral equation (55) corresponding to a CNT of length $L = 150 nm$ and $\mu = 0.1 eV, f = 2 Thz$. Real part of kernel (left) and Imagenary part (right).

The resulting equation (55) for the current density is a Fredholm equation of the second kind, which generally has a unique solution. The kernel (56) is not symmetric (the real part is centro-symmetric while the imaginary part is symmetric). It is not convenient for all types of calculations because it requires twofold numerical integration (over the azimuthal angle in (51) as well as over the longitudinal axis in (56)). However, it is possible to transform this integral expression into a more convenient alternate form, by using for the Green function (51) the corresponding cylindrical mode expansion [78]. Further analytical details are providing in Appendix C and some plots the kernel behavior are depicted in Fig 3.

## VI. NUMERICAL SIMULATIONS AND ANALYSIS

In this section, we present some numerical results from the simulations of scattering by CNT over a broad frequency spectrum, spanning from the terahertz (THz) to the infrared range. Recent technological advancements have enabled the fabrication of CNTs with a wide range of lengths [76], [77], from ultra-short tubes (~10 nm) to exceptionally long ones (0.7–10 mm). The results below encompass this broad, yet experimentally attainable, length scale. For the low-frequency regime, the scattering behavior can be characterized primarily in terms of the polarizability.

$$\alpha_{zz}^{eff} = \frac{2\pi i R_{CN}}{\omega \varepsilon_0} \int_{-L}^{L} j_z(z) dz \qquad (60)$$

and for high frequencies, we will consider the following normalized scattering pattern:

$$F(\theta) = \frac{2\pi i R_{CN}}{\omega \varepsilon_0} \sin(\theta) \int_{-L}^{L} j_z(z) e^{-ikz\cos(\theta)} dz \qquad (61)$$

At Fig. 4 we present the behavior of real and imaginary parts of nonlocality parameter $\tilde{\alpha}(\omega)$ on frequency and electrochemical potential. We don't present here the similar Figs for the local limit of conductivity because it was in detail considered in some previous papes (for example, [53]). The sharp dip at these curves corresponds to the condition $\hbar\omega \approx 2\mu$. Figure 5 presents a representative scattering profile at terahertz (THz) frequencies, derived from the numerical solution of Eq. (55). The resulting current distribution exhibits a standing-wave pattern, with a half-wavelength matching the length of the CNT. This current distribution precisely satisfies the edge condition defined by Eq. (47). The current is a complex quantity, with a dominant imaginary part. Notably, the current wavelength is much smaller than the free-space wavelength of the incident field, indicating the plasmonic nature of the excitation and a pronounced phase delay. The corresponding scattering pattern (Fig. 5b) resembles that of an effective electric dipole, which is characteristic of THz-frequency responses and persists even when spatial dispersion is taken into account.

Fig. 6 depicts the polarizability, which characterizes the induced dipole moment of the CNT in the low-frequency regime. As observed, both the real and imaginary components of the polarizability diminish with increasing frequency. This physically expected trend serves as a validation of the high accuracy achieved in the numerical computations. Notably, the presence of a resonance peak in the polarizability at low frequencies signifies the excitation of a fundamental mode, wherein the CNT length is much smaller than the wavelength of the incident field. This resonance is a key feature of the dipolar response and highlights the efficient coupling between scale the external field and the CNT at sub-wavelengths.

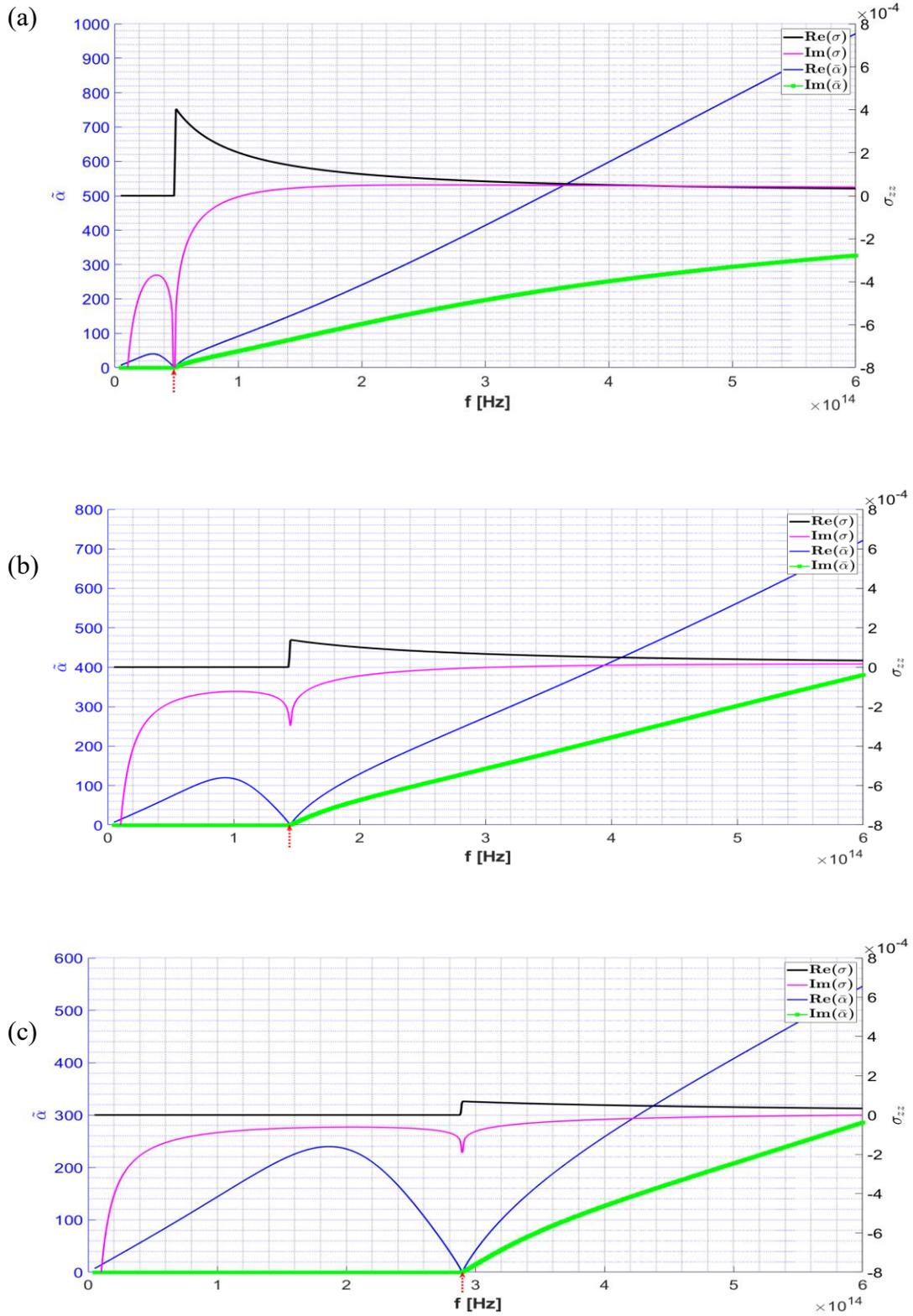

FIG.4. Plot of $\tilde{\alpha}$ as a function of frequency for a (12,0) zigzag CNT with $R_{CN} = 0.47\text{nm}, L = 216\text{nm}$. Red arrows correspond to the critical frequency $f_{Cr}$ : (a) $\mu = 0.1\text{eV}, f_{Cr} = 0.474 \cdot 10^{14}\text{Hz}$ (b) $\mu = 0.3\text{eV}, f_{Cr} = 1.450 \cdot 10^{14}\text{Hz}$ (c) $\mu = 0.6\text{eV}, f_{Cr} = 2.898 \cdot 10^{14}\text{Hz}$.

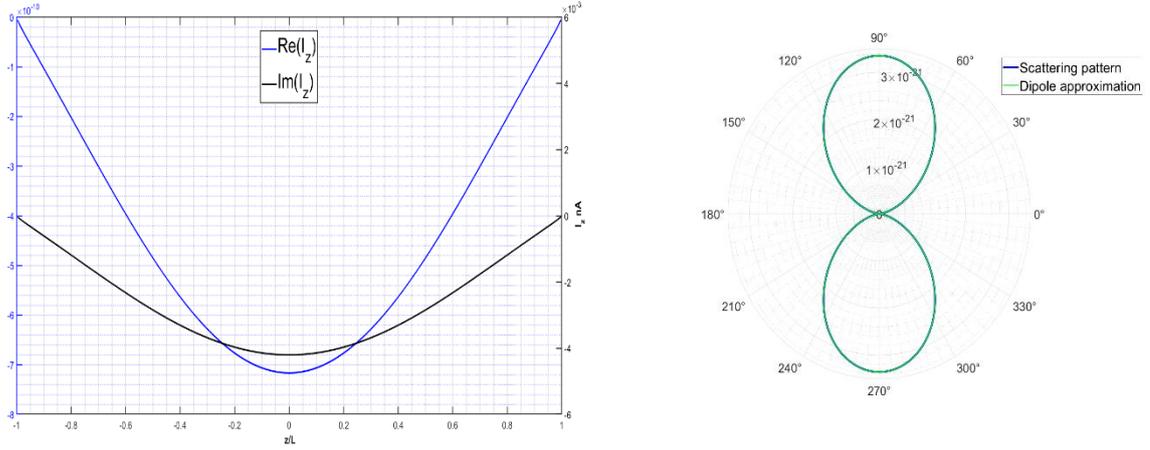

FIG. 5. Current density distribution (left) and scattering pattern (right) for a zigzag (12,0) CNT with $R_{CN}$ =0.4697nm, $L$=150nm, $f$=5 THz and $\mu$=0.1$eV$. The number of segments along the CNT used in the numerical solution of the integral equation is $N$=261.

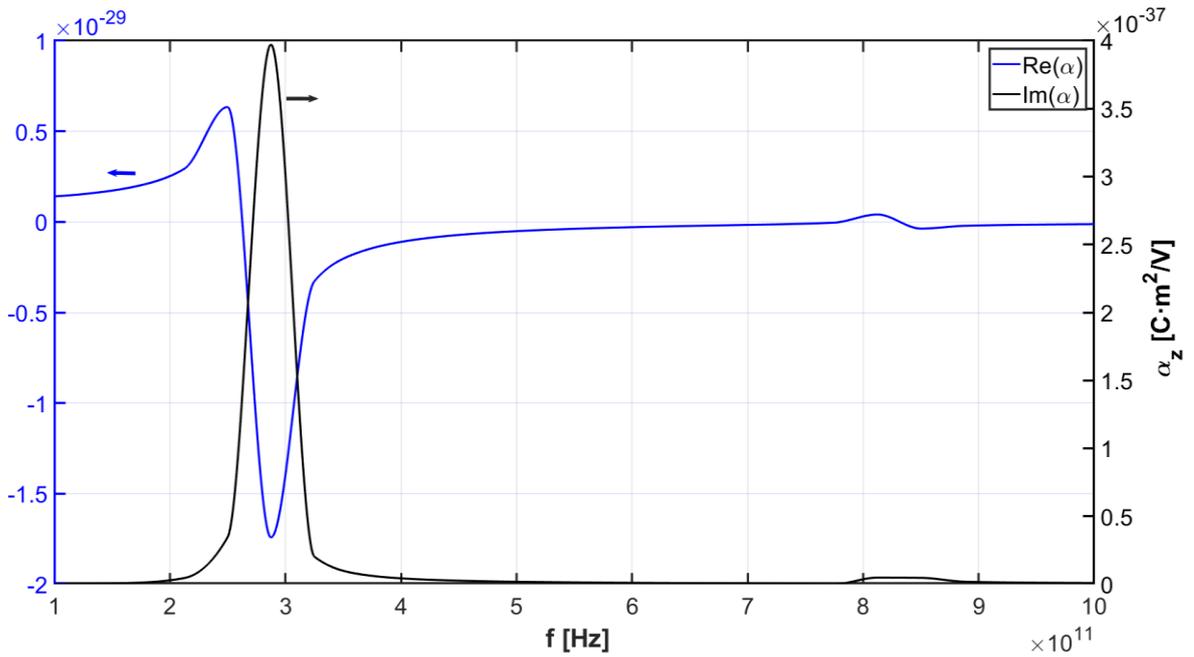

FIG.6. Polarizability as a function of frequency, resulting from accounting for nonlocality, taking $L$=1000nm, $T$=300K, $\mu$=0.1eV and $R_{CN}$ =0.47nm.

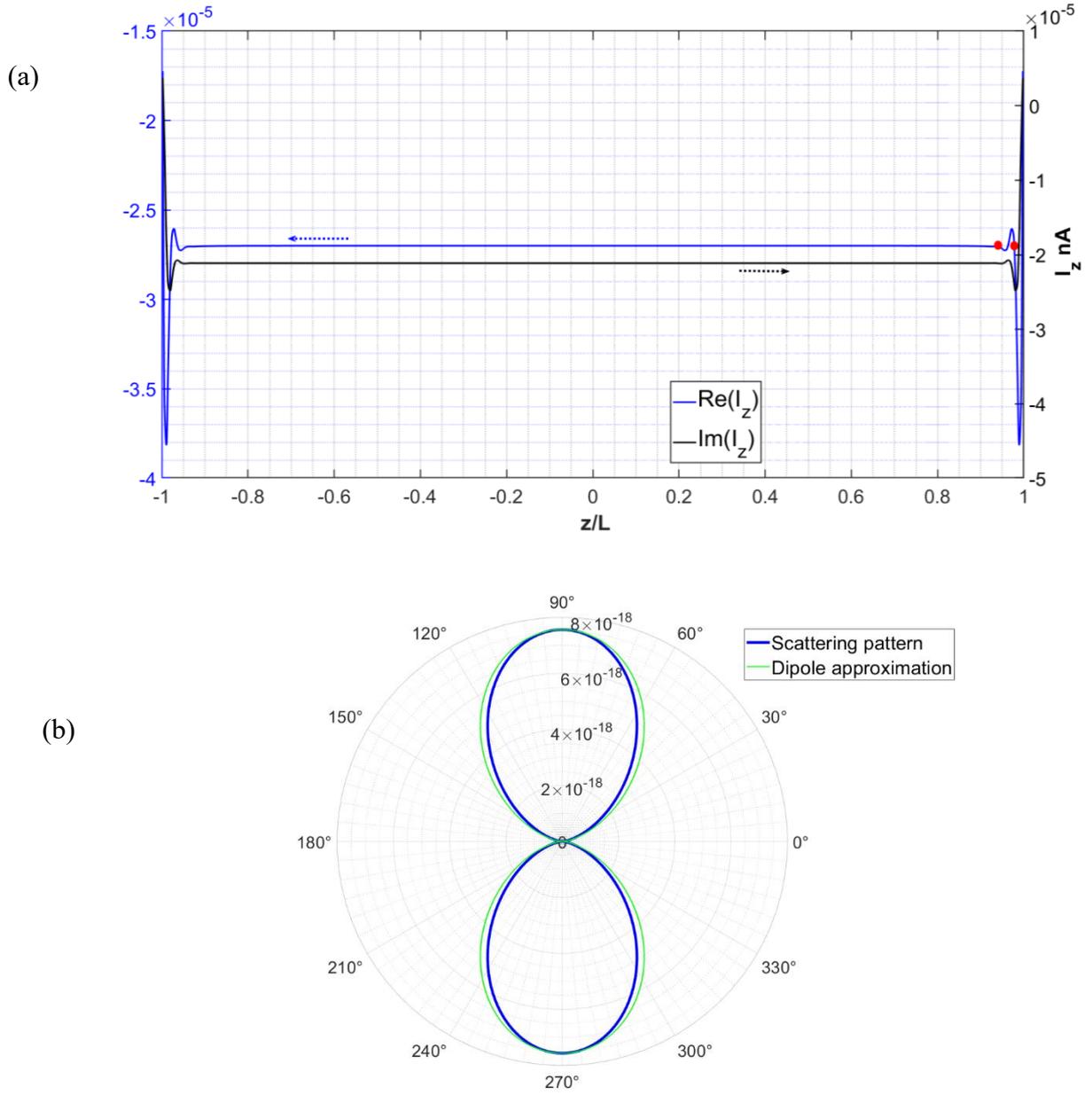

FIG. 7. A (12,0) zigzag CNT with, $R_{CN}$ =0.47nm, $L$=150nm , $\mu$=0.1$eV$ and $T$=300K is impinged in normal incidence by a plane wave of amplitude of 1v/m and $f$=300 THz.(a) Current density distribution, The number of segments taken along the CNT used in the numerical solution of the integral equation is $N$=461.The period of oscillations is defined as the distance between two neighboring maxima of the current distribution (colored red points). This value is found to be 0.0358, while $2\pi/Re(\tilde{\alpha}(\omega)L) = 0.0219$. (b) Scattering pattern.

The current density on the CNT surface at infrared frequencies, obtained through the numerical solution of Eq. (55), is illustrated in Fig. 7. The constitutive parameters used for the CNT were computed via Eqs. (20) and (21), where the interband electronic transitions dominate the response at these frequencies. As shown, the real and imaginary components of the current density exhibit similar qualitative behavior and are of comparable magnitude. Over the central portion of the CNT, the current density remains nearly uniform, while in the narrow regions near the CNT ends, it rapidly decays to zero, satisfying the physical boundary conditions. A prominent feature of the numerical simulations is the appearance of strong oscillations in the current near the ends, which attenuate quickly with increasing distance from these regions. These oscillations are a direct manifestation of nonlocal effects and underscore the significance of spatial dispersion in accurately capturing the physical response of finite-length CNTs. Specifically, the characteristic period of the oscillations, defined as the spacing between adjacent maxima in the current distribution (highlighted by red points in Fig. 7), can be quantitatively extracted from the numerical data with a value of 0.023, while the analytical value is given by $2\pi/Re(\tilde{\alpha}(\omega)) = 0.018$. The manifestation of non-locality in the current density distribution may be considered as one of the key results of this paper.

The physical interpretation of this non-local phenomenon can also be drawn from the perspective of physical optics, specifically through Huygens principle [77], [78]. However, unlike traditional applications of the principle in EM field diffraction problems, we apply it here for the motion of conductive electrons in a finite-length CNT. In classical optics, Huygens principle is often illustrated by the normal incidence of a plane wave onto a screen with a slit. In the incident half-space, the wave amplitude is uniform; however, transmission through the slit produces a central region of transparency that is flanked by shadow zones in the transmitted half-space. Geometrical optics predicts the possible occurrence of sharp discontinuities at the boundaries of these zones. Huygens principle resolves these discontinuities by introducing secondary wave sources along the slit, smoothing out the wavefronts.

In a similar manner to the case of electron motion on a CNT, an infinitely long nanotube allows unimpeded propagation of current along its axis, analogous to free-space wave propagation. When the CNT has finite length, its ends act like the edges of the slit in optical analogies, namely boundaries beyond which electrons cannot propagate. These ends give rise also to localized charge accumulations, akin to secondary sources in the Huygens framework. These point-like charges generate additional current components that resemble diffracted waves. Such edge-induced currents are localized near the CNT terminations and decay oscillatory along the CNT axis, as observed in our numerical simulations. Consequently, the total current distribution tends toward a uniform profile sufficiently far from the ends, closely paralleling the smoothing effects seen in physical diffraction theory. The period of the current oscillations near the ends is equal to $2\pi/\mathrm{Re}(\tilde{\alpha}(\omega))$, while in diffraction it is equal to $2\pi/k$. This qualitative scenario remarkably agrees with a simple analytical solution. The last one may be obtained using iterative approach and confine oneself to zero iteration.

By omitting the integral contribution in Eq. (55), we arrive at an approximate solution $j^{(0)}(z) \approx -B(z)$ whereby integrating (53), one gets

$$j^{(0)}(z) \approx \sigma_{zz}(\omega) E_z^{inc} \left(1 + \frac{\sin(\tilde{\alpha}(\omega)(z-L))}{\sin(2\tilde{\alpha}(\omega)L)} - \frac{\sin(\tilde{\alpha}(\omega)(z+L))}{\sin(2\tilde{\alpha}(\omega)L)}\right) \quad (62)$$

Such an approximation satisfies the correct CNT end condition. In the vicinity of the ends, it exhibits a linear dependence $j^{(0)}(z) \approx O(z \pm L)$, which agrees with the exact behavior of Eq. (47), obtained from WH solution. In the sequel, we name this solution as "the incident field approximation". The first term in (62) has a form of dynamical Ohm's law without nonlocality.

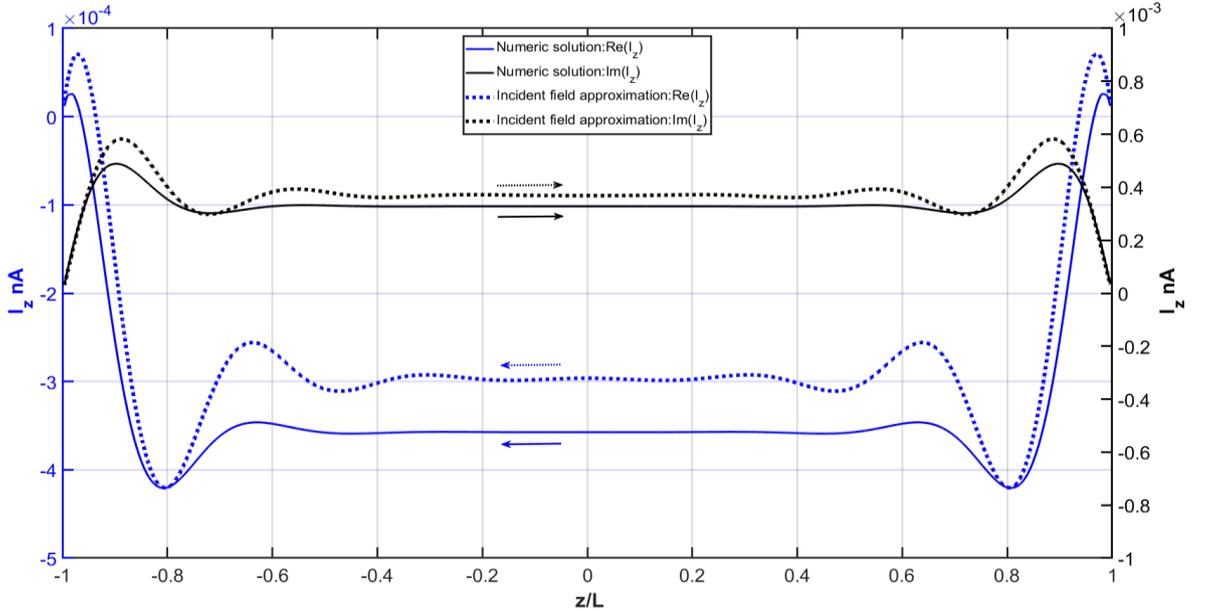

FIG 8. The incident field approximation and numerical solution for the real and imaginary parts of the current density for zigzag (12,0) CNT with $R_{CN}$ =0.47nm, $L$=150nm, $f$=200Thz, µ=0.4eV. Number of segments N=161.

This term has a finite value at the ends, which looks like the boundary light-shadow in geometrical optics. The two other terms make the current continuous and oscillating. They play the role of the previously mentioned secondary point charges. As can be seen in Fig. 8, the analytical solution agrees with the numerical simulations.

## VII. ROLE OF NONLOCALITY IN OPTOMECHANICAL FORCES

One of the prominent areas of contemporary research is the study of mechanical interactions between nanostructures mediated by electromagnetic forces—such as radiation pressure, gradient forces, and Casimir forces. A foundational contribution to this field was made by Ashkin through his pioneering development of optical tweezers, which rely on the concept of the gradient force. For Rayleigh dielectric particles, i.e., particles which are significantly smaller than the wavelength of the incident light, the gradient force is directly proportional to the spatial gradient of the incident field intensity. In the case of a uniform plane wave, this gradient clearly vanishes, resulting in a zero net force. Typically, the incident field in optical trapping experiments is a tightly focused beam, with characteristic dimensions ranging from tens of nanometers to several micrometers [81], [82]. As demonstrated in [82], when the size of the particle becomes comparable to the wavelength, the gradient force must instead be evaluated using the total field intensity, which includes both the incident and scattered field components. This refinement enables novel phenomena such as optical pulling forces, where particles are drawn toward the source rather than pushed away. The standard method for calculating such optomechanical forces involves integrating the Maxwell stress tensor (MST) over the surface enclosing the nanostructure [81], [82]. However, a more realistic representation of the system necessitates the use of finite-length models, which may introduce some fundamental challenges, particularly in accurately capturing the contributions from edge (end) effects. For CNTs, the optomechanical force is generally obtained by integrating the MST across the entire CNT surface including both axial ends, i.e.,

$$\mathbf{F} = \int_S \ddot{\mathbf{T}} \cdot \mathbf{n} dS = 2\pi R_{CN} \int_{-L}^{L} \left( \ddot{\mathbf{T}}^{(+)} - \ddot{\mathbf{T}}^{(-)} \right) \mathbf{e}_r dz \tag{63}$$

where $\ddot{\mathbf{T}}^{(\pm)}$ are the values of the stress tensor on both sides of the CNT, which can be expressed as

$$\ddot{\mathbf{T}}^{(\pm)} = \frac{1}{2} \text{Re} \left[ \left( \varepsilon_0 \mathbf{E}^{(\pm)} \otimes \left( \mathbf{E}^{(\pm)} \right)^* + \mu_0 \mathbf{H}^{(\pm)} \otimes \mathbf{H}^* \right) - \frac{1}{2} \left( \varepsilon_0 \left| \mathbf{E}^{(\pm)} \right|^2 + \mu_0 \left| \mathbf{H}^{(\pm)} \right|^2 \right) \mathbf{I} \right] =$$

$$\frac{1}{4} \text{Re} \begin{pmatrix} \varepsilon_0 \left| E_r^{(\pm)} \right|^2 - \varepsilon_0 \left| E_z^{(\pm)} \right|^2 - \mu_0 \left| H_\varphi^{(\pm)} \right|^2 & 0 & 2\varepsilon_0 E_r^{(\pm)} \left( E_z^{(\pm)} \right)^* \\ 0 & \mu_0 \left| H_\varphi^{(\pm)} \right|^2 - \varepsilon_0 \left| E_r^{(\pm)} \right|^2 - \varepsilon_0 \left| E_z^{(\pm)} \right|^2 & 0 \\ 2\varepsilon_0 E_z^{(\pm)} \left( E_r^{(\pm)} \right)^* & 0 & \varepsilon_0 \left| E_z^{(\pm)} \right|^2 - \varepsilon_0 \left| E_r^{(\pm)} \right|^2 - \mu_0 \left| H_\varphi^{(\pm)} \right|^2 \end{pmatrix} \tag{64}$$

where **n** is the outward unit vector, $\mathbf{e}_r$ is the radial unit vector and $\otimes$ represents a tensor product. It is worth noting that the local model yields unphysical results, as the integration of radial component of the electric field in Eq. (64) diverges due to non-integrable singularities in the diagonal elements. Nevertheless, by accounting for nonlocal effects these singularities are eliminated and as a result the integration involved in Eq. (64) becomes convergent. Consequently, incorporating spatial dispersion effects, provides a pathway to obtain a physically accurate and mathematically well-posed description for optomechanical interactions in carbon nanotubes (CNTs). Using the continuity of $E_z$ at the CNT surface and following Eq. (63), we obtain for the longitudinal optomechanical force

$$F_z = 2\pi\varepsilon_0 R_{CN} \cdot \mathrm{Re} \int_{-L}^{L} \left[ (E_r E_z^*)^{(+)} - (E_r E_z^*)^{(-)} \right] dz = 2\pi\varepsilon_0 R_{CN} \cdot \mathrm{Re} \int_{-L}^{L} \left( E_r^{(+)} - E_r^{(-)} \right) E_z^* dz \qquad (65)$$

The discontinuity in the radial electric field is proportional to the surface charge density. Applying the continuity relation in conjunction with Eq. (65), we find that $F_z = \frac{\pi R_{CN}}{i\omega} \int_{-L}^{L} \left( \frac{\partial j_z}{\partial z} E_z^* - \frac{\partial j_z^*}{\partial z} E_z \right) dz$.

Integrating by parts and accounting (47), we obtain

$$F_z = i\frac{\pi R_{CN}}{\omega} \int_{-L}^{L} \left( \frac{\partial E_z^*}{\partial z} j_z - \frac{\partial E_z}{\partial z} j_z^* \right) dz . \qquad (66)$$

Equation (66) is the final result, which is used for the calculation of the optomechanical force using the numerical technique developed in this paper. This relation may be applied to the derivation of an approximate analytical relation for the optomechanical force. Let us next consider a CNT in the local case, letting $j_z = \mathrm{Im}(\sigma_{zz}) E_z$ and $E_z \approx E_z^{inc}$, which suggests that $F_z \approx \frac{\pi R_{CN} \cdot \mathrm{Im}(\sigma_{zz})}{2\omega} \int_{-L}^{L} \frac{\partial |E_z^{inc}|^2}{\partial z} dz$. This is the ordinary relation for the gradient force, which vanishes for a plane wave incidence under an arbitrary angle, as mentioned before. Accounting for non-locality, one can use an approximate solution, by taking $E_z \approx E_z^{inc}$ and $j_z(z) = j_z^{(0)}(z) = -B(z)$ in Eq. (66). Let us also consider the following case of an oblique incidence of plane wave $E_z^{inc}(z) = E_0 \sin\theta_0 e^{ikz\cos\theta_0}$, where $\theta_0$ denotes the angle of incidence. For simplicity, we assume a rather short, such that $e^{2ikL\cos\theta_0} \approx 1$, which following Eq. (66) results in

$$F_z = \frac{\pi L R_{CN}}{c} \sigma_{zz}(\omega) \sin^2\theta_0 \cos\theta_0 \left[ 1 + \frac{1}{\tilde{\alpha}(\omega)L} \left( \mathrm{ctg}(2\tilde{\alpha}(\omega)L) - \frac{1}{\sin(2\tilde{\alpha}(\omega)L)} \right) \right] + c.c. \qquad (67)$$

Eq. (67) demonstrates, that non-locality qualitatively changes the behavior of optomechanical force. The first term of Eq. (67) corresponds to the local limit and is independent on the non-local parameter $\tilde{\alpha}(\omega)$ and is proportional to the CNT length. Such a behavior corresponds to the Ashkin's model of optical tweezing. The second termin Eq. (67) describes the influence of non-locality. It is a non-monotonic and an alternating (oscillatory) quantity with respect to the lengthof the CNT.

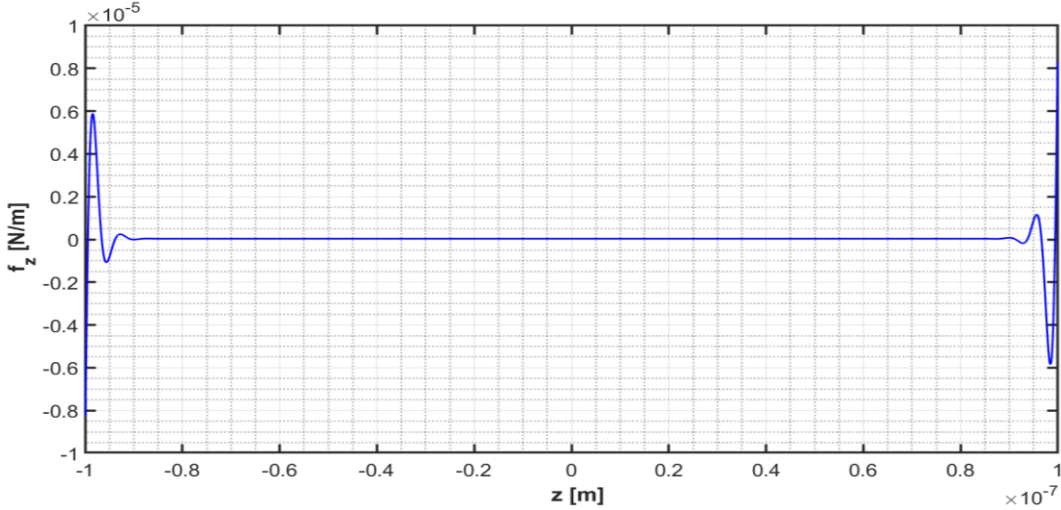

FIG.9. Force density $f_z(z)$ as a function of the longitudinal coordinate z in the case of oblique incidence ($\theta_0$=45°) for $L$=100nm, $T$=300K, $R_{CN}$=0.46973nm, $f$=200 THz. The incident field is $E_0 = 10^7$ V/m, µ=0.1eV.

Fig. 9 presents the distribution of the longitudinal force density along the CNT axis. The force exhibits pronounced oscillations near the ends, which decay rapidly at distances farther from the terminations. These oscillations resemble those observed in the current density distribution in Fig. 9 and are similarly originated from the nonlocal nature of the CNT conductivity. This spatially varying force can excite longitudinal elastic waves within the CNT through its interaction with incident light. Figure 10 displays the total longitudinal optical force as a function of CNT length for various values of electrochemical potential. At large lengths, the influence of nonlocality diminishes, and the force behavior resembles that of conventional dielectric particles, e.i, a pushing force that increases with length. The magnitude of this force is on the order of 1.0 pN, comparable to typical gradient forces encountered in optical tweezing applications [83]. However, for short CNTs with sufficiently high electrochemical potential, nonlocal effects become significant, resulting in a qualitative shift: the optical force transitions from pushing to pulling, as evidenced by the negative force values shown in Fig. 10 (b).

## VIII. CONCLUSION AND OUTLOOK

A new framework of CNT electrodynamics with non-local conductivity is provided. It is based on the numerical and analytical solution of a Fredholm integral equation and is applied in analyzing the influence of nonlocality in the scattering problem of EM-field by a finite CNT. It is shown, that a realistic scattering model of a CNT, requires considering finite-length effects. Correct formulations of integral equations, demand the account of nonlocality, while ignoring it may produce unphysical results in the numerical simulations and non-physical predictions. It is demonstrated that nonlocality creates additional modes, which are manifested by the appearance of new types of resonances. The origin of these resonances is essentially different from ordinary harmonic oscillators and are by the asymmetry of the corresponding spectral lines. Our results can be directly applied for the realization of new research directions in nano-antennas, nanocircuits and sensing based on CNTs. Another relevant field of immense potential, is CNT based technology of metamaterials and metasurfaces. The present methodology using a single CNT can be extended to also account for the scattering problem between several CNTs in close proximity as well as for describing an equivalent composite medium using a suitable type of homogenization. The obtained results are applied for the analysis of optical forces in CNTs, indicating that nonlocality considerably changes their qualitative behavior of such forces.

In summary, this paper should be regarded as a foundational contribution to the nonlocal electrodynamics of CNTs, with the aim of stimulating broader explorations in this emerging area of research. Several promising new directions for future research naturally follow from this study. Once the current density distribution in a finite CNT has been determined, the associated Joule heating problem—namely, the spatial distribution of temperature and heat generation—can also be addressed, thereby extending the analysis previously applied to finite nanowires [84] to the present case of finite-length CNTs. One particularly compelling problem, is the analysis of interactions between two or more CNTs (same or different lengths), which would lead to a system of coupled integral equations similar to Eq. (57), with similar Green's functions in the kernel. Another avenue of interest involves CNTs suspended between electrodes, where the boundary conditions must be modified in order to account for electron penetration into the contacts. This specific scenario, necessitates an additional adjustment of the Green's function in Eq. (53) to enforce the proper electromagnetic boundary conditions applied at the electrode surfaces. Some additional research opportunities include for example, investigating how nonlocal CNT properties are affected by external stimuli, such as bias voltages, magnetic fields, chemical doping, etc. Such extensions could offer further insight into the tunability of CNTs in practical nanoscale applications.

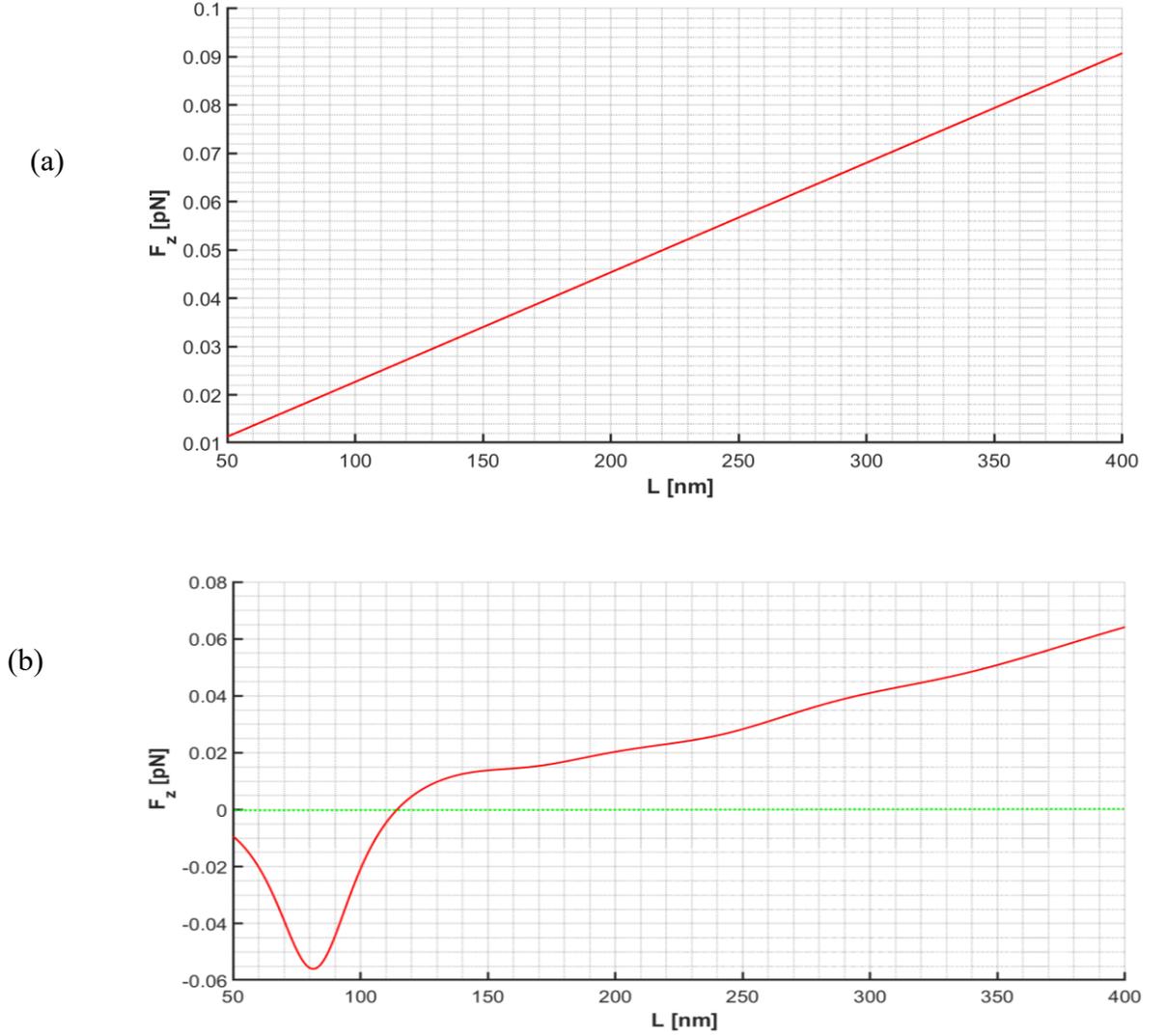

FIG. 10. Total longitudinal force $F_z$ as a function of CNT's length for different values of electrochemical potential for oblique incidence ($\theta=30°$). T=300K, $R_{CN}$ =0.47nm , f=200 THz. The amplitude of the incident field is $E_0 = 10^7$ V/m ; chemical potential is (a) 0.1eV; (b) 0.4125eV .

**APPENDIX A: WIENER-HOPF EQUATION FOR CARBON NANOTUBE**
  1. **Deduction of the Wiener-Hopf equation**

For convenience, we will perform the calculations for the complex wavenumber $k = k' + ik'', k'' > 0$ and make the transition $k'' \to 0$ in the final result. The Hertz potential is presented in the form of a Fourier integral $\Pi(r,z) = \int_{-\infty}^{\infty} \bar{\Pi}(r,\alpha) e^{-i\alpha z} d\alpha$, with the spectral amplitude $\bar{\Pi}(r,\alpha)$ found from the Helmholz equation and given by the relation (41). The current density is given as

$$i\omega\varepsilon_0\left(\frac{\partial \Pi^{(+)}(z)}{\partial r} - \frac{\partial \Pi^{(-)}(z)}{\partial r}\right) = \begin{cases} 0, z < 0 \\ j_z(z), z > 0 \end{cases} \quad (A1)$$

Expressed in terms of Fourier amplitudes, (A1) reads

$$i\omega\varepsilon_0\left(\frac{\partial \bar{\Pi}^{(+)}(\alpha)}{\partial r} - \frac{\partial \bar{\Pi}^{(-)}(\alpha)}{\partial r}\right) = \Upsilon_+(\alpha) \quad (A2)$$

where $\Upsilon_+(\alpha)$ is the Fourier transformation of the current density (symbol + means its analyticity at the top half-plane due to its vanishing for $z < 0$). Continuity of the Hertz potential along the whole axis $z$, implies that

$$A(\alpha)K_0(\gamma R_{CN}) = D(\alpha)I_0(\gamma R_{CN}) \quad (A3)$$

Next, combining (41) with (A2), renders

$$\Upsilon_+(\alpha) = -i\omega\varepsilon_0\gamma\left[A(\alpha)K_1(\gamma R_{CN}) + D(\alpha)I_1(\gamma R_{CN})\right] \quad (A4)$$

and substituting of (A3) into (A4), by excluding $D(\alpha)$ from (A4), leads to

$$\Upsilon_+(\alpha) = -i\omega\varepsilon_0\gamma A(\alpha)\left[K_1(\gamma R_{CN}) + \frac{I_1(\gamma R_{CN})}{I_0(\gamma R_{CN})}K_0(\gamma R_{CN})\right] \quad (A5)$$

Relation (A5) allows us to connect the amplitudes $A(\alpha), D(\alpha)$ with the current density. Using (A3), (A5) and the $I_0(x)K_1(x) + K_0(x)I_1(x) = x^{-1}$, yields

$$D(\alpha) = i\frac{R_{CN}}{\omega\varepsilon_0}K_0(\gamma R_{CN})\Upsilon_+(\alpha) \quad (A6)$$

$$A(\alpha) = i\frac{R_{CN}}{\omega\varepsilon_0}I_0(\gamma R_{CN})\Upsilon_+(\alpha) \quad (A7)$$

By applying a Fourier transformation to the boundary condition in Eq. (39), we finally get

$$\left(1 - \frac{\alpha^2}{\tilde{\alpha}^2}\right)\Upsilon_+(\alpha) = -\gamma^2\sigma_{zz}\left(\bar{\Pi}_+^{(+)}(\alpha) + \bar{\Pi}_-^{(+)}(\alpha)\right) + \sigma_{zz}E_0\int_0^\infty e^{i(k\cos\theta_0 + \alpha)z}dz \quad (A8)$$

Expressing the value $\bar{\Pi}_+^{(+)}(\alpha)$ through the current and preforming the integration, one obtains equation (42), which has the standard form of WH equation $\gamma^2 \Upsilon_+(\alpha) G(\alpha) = \Xi_-(\alpha) + \dfrac{\omega \varepsilon_0 E_0}{(\alpha + k\cos\theta_0) R_{CN}}$, where $\Xi_-(\alpha)$ is the second unknown function, which is analytical in the bottom half-plane of the complex variable $\alpha$. Note that the WH equation holds in a strip $\tau_- < \tau < \tau_+$ in the complex $\alpha$ plane.

### 2. Solution of Wiener-Hopf equation

The WH solution may be solved exactly in analytical form [46]. The key step is to decompose the arbitrary function $Q(\alpha)$ into two functions $Q_\pm(\alpha)$, where $Q_+(\alpha)$ is regular at the top half-plane $\tau > \tau_-$ and $Q_-(\alpha)$ is regular at the bottom half-plane $\tau < \tau_+$. Such functions are given by the following integrals

$$Q_+(\alpha) = \frac{1}{2\pi i} \int_{-\infty+ic}^{\infty+ic} \frac{Q(\beta)}{\beta - \alpha} d\beta, \tau_- < c < \tau < \tau_+ \tag{A9}$$

$$Q_-(\alpha) = -\frac{1}{2\pi i} \int_{-\infty+id}^{\infty+id} \frac{Q(\beta)}{\beta - \alpha} d\beta, \tau_- < \tau < d < \tau_+ \tag{A10}$$

In a similar manner, the arbitrary function $G(\alpha)$ which is analytical on the strip $\tau_- < \tau < \tau_+$, may be presented as a product $G(\alpha) = G_+(\alpha) \cdot G_-(\alpha)$ by first taking the logarithm, and then performing the sum decomposition. The result reads as $G_+(\alpha) = e^{W_+(\alpha)}$, $G_-(\alpha) = e^{W_-(\alpha)}$, where

$$W_+(\alpha) = \frac{1}{2\pi i} \int_{-\infty+ic}^{\infty+ic} \frac{\ln[G(\beta)]}{\beta - \alpha} d\beta, \tau_- < c < \tau < \tau_+ \tag{A11}$$

$$W_-(\alpha) = -\frac{1}{2\pi i} \int_{-\infty+id}^{\infty+id} \frac{\ln[G(\beta)]}{\beta - \alpha} d\beta, \tau_- < \tau < d < \tau_+ \tag{A12}$$

for even functions $G(\alpha)$, we clearly have $G_+(-\alpha) = G_-(\alpha)$. Using factorization of the function $G(\alpha)$, we can then rewrite the WH equation in the following form

$$(\alpha + k)\Upsilon_+(\alpha) G_+(\alpha) = \frac{\Xi_-(\alpha)}{(\alpha - k) G_-(\alpha)} + \frac{\omega \varepsilon_0 E_0}{(\alpha + k\cos\theta_0)(\alpha - k) G_-(\alpha) R_{CN}} \tag{A13}$$

which may be also presented as

$$(\alpha+k)\Upsilon_+(\alpha)G_+(\alpha) = \frac{\Xi_-(\alpha)}{(\alpha-k)G_-(\alpha)} + S_+(\alpha) + S_-(\alpha) \tag{A14}$$

where

$$S_+(\alpha) = \frac{\omega\varepsilon_0 E_0}{2\pi i} \int_{-\infty+ic}^{\infty+ic} \frac{d\beta}{(\beta+k\cos\theta_0)(\beta-k)(\beta-\alpha)G_-(\beta)} \tag{A15}$$

Note that we don't present here the corresponding expression for $S_-(\alpha)$, since it is redundant for the present analysis. The integration contour in Eq. (A15) must be closed in the upper half of the complex plane. Within this contour, the integrand possesses only a single singularity located at a specific point (i.e., a simple pole at the point $\beta = -k\cos\theta_0$). In terms of its residue, the value of the integral can be written as

$$S_+(\alpha) = \frac{\omega\varepsilon_0 E_0}{k(1+\cos\theta_0)(\alpha+k\cos\theta_0)G_+(k\cos\theta_0)} \tag{A16}$$

Equation (A14) shows the equality between the of two functions, one of which is analytical in the top half-plane, while the other is analytical in the bottom half-plane. This means, that both functions are equal to the integer function [48]. According to the Meixner concept, this function must be set to zero [48] and as a result we find that $(\alpha+k)\Upsilon_+(\alpha)G_+(\alpha) = S_+(\alpha)$. Finally,

$$\Upsilon_+(\alpha) = \frac{\omega\varepsilon_0 E_0}{k(1+\cos\theta_0)(\alpha+k\cos\theta_0)G_+(k\cos\theta_0)G_+(\alpha)(\alpha+k)} \tag{A17}$$

The integrals (A11),(A12) for the function $G(\alpha)$ can not be evaluated in an analytical form, however their asymptotic behavior at $\alpha \to \infty$, can still be determined analytically. By inverting the order of the Fourier transform in (A17), we obtain Eq. (44) for the current density and Eq. (45) for the charge density. The final results for the Hertz potential and electric field components, are given below as

$$\Pi(r,z) = \frac{i\eta}{k} \int_C e^{-i\alpha z} \frac{I_0(\gamma R_{CN})K_0(\gamma r)}{(\alpha+k\cos\theta_0)(\alpha+k)G_+(\alpha)} d\alpha \tag{A18}$$

$$E_z^{sc}(r,z) = -\frac{i\eta}{k}\int_C e^{-i\alpha z}\frac{(\alpha-k)I_0(\gamma R_{CN})K_0(\gamma r)}{(\alpha+k\cos\theta_0)G_+(\alpha)}d\alpha \qquad (A19)$$

$$E_r^{sc}(r,z) = -\frac{\eta}{k}\int_C e^{-i\alpha z}\alpha\sqrt{\frac{\alpha-k}{\alpha+k}}\frac{I_0(\gamma R_{CN})K_1(\gamma r)}{(\alpha+k\cos\theta_0)G_+(\alpha)}d\alpha \qquad (A20)$$

The relations (A18) - (A20) are valid in the exterior region $r > R_{CN}$. The corresponding expressions prevailing in the interior region $r < R_{CN}$, may be obtained via simple exchange of $r \Leftrightarrow R_{CN}$.

**APPENDIX B: DERIVATION OF THE RENORMALIZED INTEGRAL IN EQ. (57)**

Let us consider the configuration of the CNT depicted at Fig.1. The symbol $\Sigma$ denotes the surface of CNT ($d\Sigma = 2\pi R_{CN}dzd\phi$ is the differential of this surface). The symbol $\Sigma_\delta$ is its exclusion element, which simplifies the transformation of singular integrals (it allows them to exclude the singularities and goes to zero at the final step of calculations). The symbols $C, C_\delta$ represent the two-side closed rims of the CNT and the excluded element, respectively. Following [61], we consider the integral

$$v(z') = \lim_{\delta\to 0}\int_{\Sigma-\Sigma_\delta} j_z(s)\left(\frac{e^{ikR}}{R}\right)d\Sigma \qquad (B1)$$

where $R = \sqrt{4R_{CN}^2\sin^2\left(\frac{\phi}{2}\right)+(s-z')^2}$.

Thus,

$$\begin{aligned}\frac{\partial v(z')}{\partial z'} &= \lim_{\delta\to 0}\int_{\Sigma-\Sigma_\delta} j_z(s)\frac{\partial}{\partial z'}\left(\frac{e^{ikR}}{R}\right)d\Sigma = \\ &-\lim_{\delta\to 0}\int_{\Sigma-\Sigma_\delta} j_z(s)\frac{\partial}{\partial s}\left(\frac{e^{ikR}}{R}\right)d\Sigma = \\ &\lim_{\delta\to 0}\left[\int_{\Sigma-\Sigma_\delta}\frac{\partial j_z(s)}{\partial s}\left(\frac{e^{ikR}}{R}\right)d\Sigma - \int_{\Sigma-\Sigma_\delta}\frac{\partial}{\partial s}\left(j_z(s)\frac{e^{ikR}}{R}\right)d\Sigma\right] = \\ &\lim_{\delta\to 0}\left[\int_{\Sigma-\Sigma_\delta}\frac{\partial j_z(s)}{\partial s}\left(\frac{e^{ikR}}{R}\right)d\Sigma - \int_C j_z(s)\frac{e^{ikR}}{R}dC\right]\end{aligned} \qquad (B2)$$

where the integral over $C_\delta$ in the last term line is identically null vanishes. Performing the second derivative gives,

$$\frac{\partial^2 v(z')}{\partial z'^2} = \lim_{\delta \to 0}\left[\int_{\Sigma-\Sigma_\delta} \frac{\partial(j_z(s)-j_z(z'))}{\partial z'}\frac{\partial}{\partial z'}\left(\frac{e^{ikR}}{R}\right)d\Sigma - \int_C j_z(s)\frac{\partial}{\partial z'}\left(\frac{e^{ikR}}{R}\right)dC\right] =$$
$$\lim_{\delta \to 0}\left[\int_{\Sigma-\Sigma_\delta} \frac{\partial(j_z(s)-j_z(z'))}{\partial z'}\frac{\partial^2}{\partial z'^2}\left(\frac{e^{ikR}}{R}\right)d\Sigma + \int_C (j_z(s)-j_z(z'))\frac{\partial}{\partial z'}\left(\frac{e^{ikR}}{R}\right)dC\right]$$
(B3)

Next, using

$$\lim_{\delta \to 0}\int_{\Sigma-\Sigma_\delta}(j_z(s)-j_z(z'))\frac{\partial^2}{\partial z'^2}\left(\frac{e^{ikR}}{R}\right)d\Sigma =$$
$$V.P.\int_{-L}^{L}\frac{\partial^2}{\partial s^2}\left(\int_0^{2\pi}\frac{e^{ikR}}{R}d\phi\right)(j_z(s)-j_z(z'))dz'$$
(B4)

and

$$\int_C (j_z(s)-j_z(z'))\frac{\partial}{\partial z'}\left(\frac{e^{ikR}}{R}\right)dC =$$
$$j_z(z')\left(\int_0^{2\pi}\frac{e^{ikR}}{R}d\phi\bigg|_{z'=-L} - \int_0^{2\pi}\frac{e^{ikR}}{R}d\phi\bigg|_{z'=L}\right)$$
(B5)

leads to

$$\frac{\partial^2 v(z')}{\partial z'^2} = V.P.\int_{-L}^{L}\frac{\partial^2 G(z'-s)}{\partial s^2}(j_z(s)-j_z(z'))ds +$$
$$j_z(z')\frac{\partial}{\partial z'}(G(z'-L)-G(z'+L))$$
(B6)

Multiplying (B6) by $g(z,z')$ and integrating over the CNT surface, one gets

$$\int_{-L}^{L}\frac{\partial^2 v(z')}{\partial z'^2}g(z,z')dz' = V.P.\int_{-L}^{L} j_z(s)\int_{-L}^{L}\frac{\partial^2 G(z'-s)}{\partial s^2}g(z,z')dz'ds -$$
$$-V.P.\int_{-L}^{L} j_z(z')\int_{-L}^{L}\frac{\partial^2 G(z'-s)}{\partial s^2}g(z,z')dsdz' + \int_{-L}^{L} j_z(z')\frac{\partial}{\partial z'}(G(z'-L)-G(z'+L))dz'$$
(B7)

Note that in the second term on the right hand side of (B7), we have exchanged $\partial^2/\partial s^2 \to \partial^2/\partial z'^2$, $s \to z', z' \to s$, and $z' \to s$ in the third term, resulting finally in Eq. (57).

## APPENDIX C: ALTERNATE FORM OF THE KERNEL IN EQ. (55) USING A CYLINDRICAL WAVE EXPANSION

Expansion the Green function of the Helmholtz equation [78] by cylindrical modes, we arrive at

$$\frac{e^{ik\sqrt{r^2+R_{CN}^2-2rR_{CN}\sin(\phi)+(s-z)^2}}}{\sqrt{r^2+R_{CN}^2-2rR_{CN}\sin(\phi)+(s-z)^2}} = \frac{i}{2}\int_{-\infty}^{\infty} e^{ih(z-s)} H_0^{(1)}\left(\sqrt{k^2-h^2}\,\bar{r}\right) dh \tag{C1}$$

where $\bar{r} = \sqrt{r^2+R_{CN}^2-2rR_{CN}\sin(\phi)}$. Using the additional theorem for the Hankel functions [78], implies that

$$H_0^{(1)}(\kappa\bar{r}) = \sum_n H_{-n}^{(1)}(\kappa r) J_n(\kappa R_{CN}) e^{in\phi} \tag{C2}$$

where $r > R_{CN}$. Substituting (C1) and (C2) Eq. (53) and integrating over the angle $\phi$ and keeping only one term with $n = 0$, gives

$$G(z-s) = G(R_{CN}, z-s) = i\pi R_{CN} \int_{-\infty}^{\infty} e^{ih(z-s)} H_0^{(1)}(\kappa R_{CN}) J_0(\kappa R_{CN}) dh \tag{C3}$$

where $\kappa = \sqrt{k^2-h^2}$. Restriction on the sign of $\mathrm{Im}(\sqrt{k^2-h^2})$ is imposed to assure the proper decay of the Hankel function as $|h| \to \infty$.

As a next step we substitute (C3) in Eq. (58), which leads to

$$K(z,s) = i\pi R_{CN}\left(\frac{\partial^2}{\partial s^2}+k^2\right)\int_{-\infty}^{\infty} e^{ihs} H_0^{(1)}(\kappa R_{CN}) J_0(\kappa R_{CN}) \left(\int_{-L}^{L} g(z,z') e^{-ihz'} dz'\right) dh \tag{C4}$$

where $g(z,z')$ is the Green function given by Eq. (51). The integration over $z'$ is elementary

$$\int_{-L}^{L} g(z,z') e^{-ihz'} dz' = -\frac{F(h,z)}{(\tilde{\alpha}^2-h^2)\sin(2\tilde{\alpha}L)} \tag{C5}$$

where

$$F(h,z) = e^{-ihz}\sin(2\tilde{\alpha}L) + e^{ihL}\sin X - e^{-ihL}\sin Y \tag{C6}$$

and $X = \tilde{\alpha}(z-L), Y = \tilde{\alpha}(z+L)$. The function (C5) doesn't have a pole singularity at $h = \pm\tilde{\alpha}$ since $F(\pm\tilde{\alpha}, z) = 0$. This function satisfies the symmetry relation $F(h,z) = F(-h,-z)$, therefore $K(z,s) = K(-z,-s)$. The order of integration over $h$ and differentiation with respect to $s$ in (C4) may be exchanged. By substituting (C5) to (C4), we obtain

$$K(z,s) = -\frac{i\pi R_{CN}}{\sin(2\tilde{\alpha}L)} \int_{-\infty}^{\infty} e^{ihs} (k^2 - h^2) \frac{H_0^{(1)}(\kappa R_{CN}) J_0(\kappa R_{CN})}{\kappa^2 - k^2} F(z,h) dh \quad (C.7)$$

This is the final result of this Appendix, which gives the kernel (56) in the form of a one-fold integral.

---

.